# Antisymmetric thermopower without magnetism in three-phase composites

Joohan Nam[1]*

[1]Chair of Inorganic Chemistry I, Technische Universität Dresden, Bergstraße 66, 01069 Dresden, Germany.

*e-mail: joohan.nam@mailbox.tu-dresden.de

**A thermopower that turns a temperature gradient into a perpendicular electric field in every in-plane direction has relied on magnetism. Reciprocity permits it without magnetism, yet under a three-fold or higher axis it vanishes both in reciprocal elastic band transport and in composites of two isotropic phases with ideal interfaces. Here we show that composites of three isotropic phases carry it, and that under such an axis three is the minimum. A third, linearly independent transport matrix escapes the exact two-phase reduction; at weak contrast the first allowed axial term is cubic, proportional to a determinant of the three matrices times material-weighted, handed three-point correlations. Treating the phases as colours identifies the arrangements that allow the response, and crystal nets from reticular chemistry supply printable three-dimensional ones. Certified two-sided bounds give axial Seebeck coefficients of 49.5–50.2 μV $K^{-1}$ in a planar cell and 76–84 μV $K^{-1}$ in a printable woodpile; reflections in planes containing the axis reverse the response, and recolouring 24 nets to restore their parent symmetry switches it off. Around a line heat source the axial driving field has the 1/r azimuthal form of the magnetic field of a straight wire and drives circulating currents.**

Thermopowers that turn a temperature gradient into a perpendicular electric field in every in-plane direction have relied on magnetism, as in the anomalous Nernst effect.[1,2] For electrical conduction this is a theorem: in time-reversal-symmetric matter, Onsager reciprocity forbids an antisymmetric conductivity.[3,4] A magnet-free antisymmetric thermopower would sense heat flux in any direction without arrays of junctions,[5,6] but passive routes have not provided one. Tilted laminates keep a mirror plane and give only a symmetric off-diagonal thermopower,[7–9] and geometric metamaterials have reshaped antisymmetric transport through the Hall effect in a magnetic field.[10–14]

For coupled transport, reciprocity constrains less. Onsager–Casimir relations bind the Seebeck tensor to the transpose of the Peltier tensor, not to itself,[15,16] so a time-even antisymmetric part is allowed in the 13 pyroaxial point groups, which lack mirror planes containing the axis and two-fold axes perpendicular to it.[16–18] In crystals it is expected usually only through misaligned principal axes in monoclinic and triclinic lattices.[16] Under a three-fold or higher axis, which makes the in-plane conductivity isotropic, reciprocal band transport with elastic scattering forbids it: the conductivity and the thermoelectric coupling tensor are moments of one symmetric transport distribution, so both are scalar in the plane (Supplementary Note 9). In composites, charge and heat can follow different paths and homogenization can break the symmetry of the Seebeck tensor,[19] but two isotropic phases with ideal interfaces reduce exactly to two scalar conduction problems,[20,21] their electrical and coupling tensors stay symmetric[22], and under that axis the axial thermopower vanishes. Whether any composite of isotropic materials can carry an antisymmetric thermopower while its in-plane conduction stays isotropic has therefore remained open.

Here we show that composites of three isotropic phases can, and that under a three-fold or higher axis three is the minimum for isotropic phases with ideal interfaces. A third phase whose transport matrix is linearly independent of the other two defeats the exact two-phase reduction; a handed arrangement then yields an antisymmetric coupling tensor, which no symmetric band-transport distribution can produce (Fig. 1). About one phase, the first allowed axial term is cubic in the contrast: the phase-count determinant times material-weighted, handed three-point correlations.[23] To find such composites in three dimensions we treat the phases as colours, as for the conductivity of planar multiphase checkerboards,[24] and use the colour symmetry that keeps every phase in place to decide which arrangements allow the axial response; we adapt the topological net library of reticular chemistry[25] as a design space for printable multimaterial lattices. Two-sided variational bounds certify 76–84 μV $K^{-1}$ in a printable segmented woodpile;[26,27] reflections in planes containing the axis reverse the sign, and recolouring 24 nets to restore their parent symmetry switches the response off. Around a line heat source the axial driving field has the 1/r azimuthal form of the magnetic field of a straight wire and drives currents in closed rings.

**Two phases cancel; three can respond**

We consider periodic composites of isotropic phases, each described by an electrical conductivity σ, a Seebeck coefficient S and a thermal conductivity κ. Linear transport in each phase is $\mathbf{j} = -\sigma(\nabla V + S\nabla T)$, and the heat flux is $\mathbf{q} = TS\mathbf{j} - \kappa\nabla T$. Homogenization gives effective tensors $\boldsymbol{\sigma}_{eff}$ and $\mathbf{L}_{eff} = (\sigma S)_{eff}$, the effective Seebeck tensor $\mathbf{S}_{eff} = \boldsymbol{\sigma}_{eff}^{-1}\mathbf{L}_{eff}$ and the open-circuit thermal conductivity $\boldsymbol{\kappa}_{eff}$ (Methods). The antisymmetric part of $\mathbf{S}_{eff}$ defines the axial coefficient $S_a = (S_{xy} - S_{yx})/2$, which adds the transverse field $\mathbf{E}_a = S_a\, \nabla T \times \hat{z} = -S_a\, \hat{z} \times \nabla T$ to the open-circuit response.

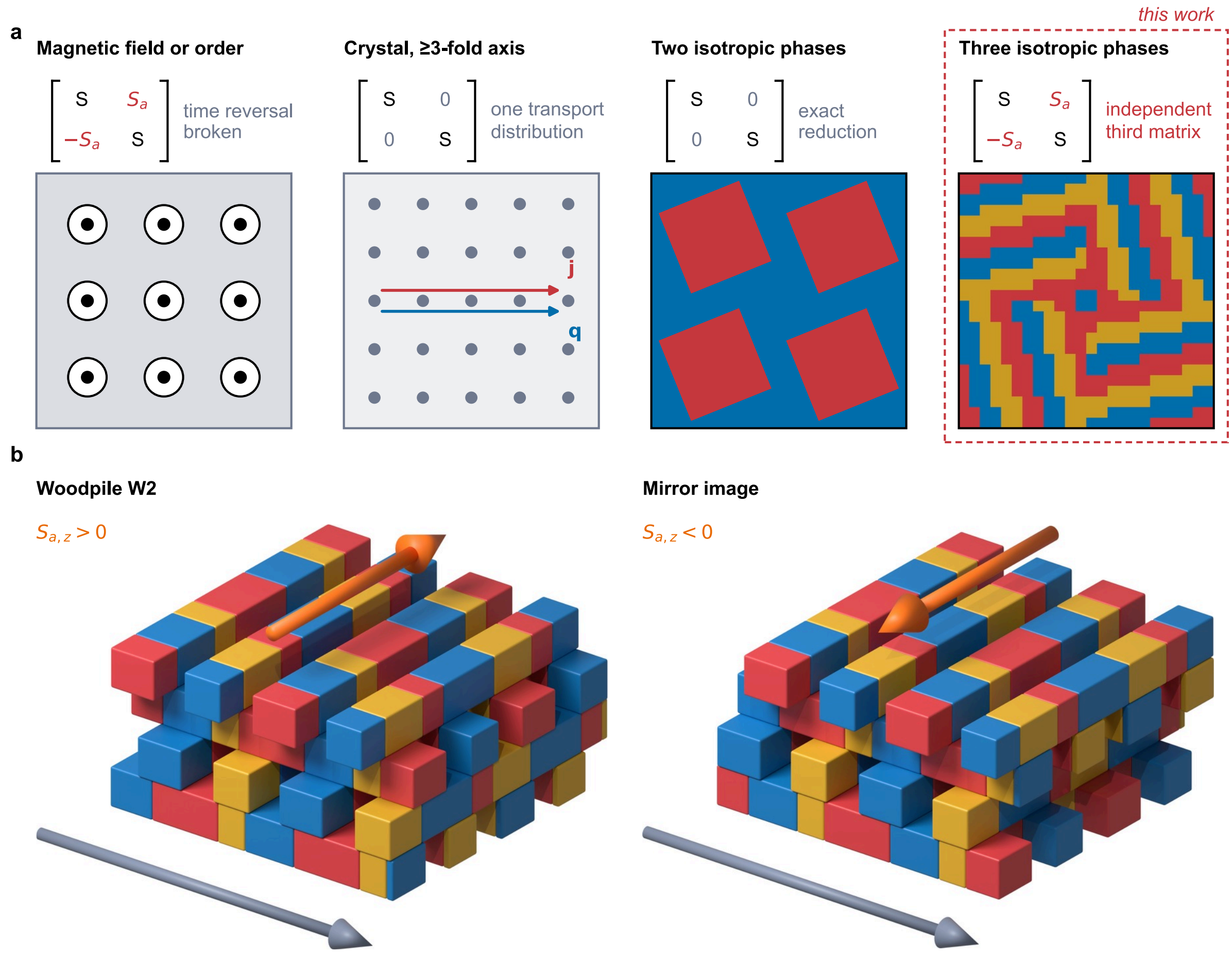

**Fig. 1 | Routes to an antisymmetric thermopower and a printable mirror pair. a**, Each box shows an arrangement and, above it, the form of its in-plane Seebeck tensor. A magnetic field or magnetic order can give an antisymmetric part by breaking time-reversal symmetry. In a crystal with a three-fold or higher axis, reciprocal elastic band transport makes the electrical and thermoelectric conductivities moments of one symmetric transport distribution, so the in-plane Seebeck tensor is scalar (Supplementary Note 9); the arrows mark the charge current **j** and the heat current **q**. Two isotropic phases reduce exactly to two scalar conduction problems, so even a handed arrangement (p-type squares, red, in an n-type matrix, blue; schematic) gives no antisymmetric part under such an axis. A third isotropic phase with an independent transport matrix (ochre) removes this reduction, and a ferroaxial-type arrangement can give an antisymmetric part in a passive, reciprocal medium. **b**, The segmented-rod woodpile W2 of Fig. 5 (2 × 2 cells; p-type red, n-type blue, weak rod segments ochre; the weak matrix, of the same material, is not drawn) and its mirror image. The same heat flow **q** (grey arrows) produces transverse fields $\mathbf{E}_a = S_{a,z}\nabla T \times \hat{z}$, with $S_{a,z} = (S_{xy} - S_{yx})/2$ along z (orange arrows), of opposite sign.

The axial coefficient is even under time reversal, so it requires no magnetism. A time-even axial vector is allowed only in the 13 pyroaxial point groups, the symmetry classes of ferroaxial domain states, which lack mirror planes containing the axis and two-fold axes perpendicular to it.[17,18] In two dimensions these are the plane groups without mirror or glide lines, *p*1, *p*2, *p*3, *p*4 and *p*6. We call an arrangement with this symmetry ferroaxial-type; its mirror image plays the part of the opposite domain. The term implies no spontaneous ferroaxial order. Reciprocity keeps $\boldsymbol{\sigma}_{eff}$ and $\boldsymbol{\kappa}_{eff}$ symmetric even in these groups.[3,4] Symmetry alone, however, is not sufficient. With two phases, the two local transport matrices can be diagonalized simultaneously, which splits the coupled problem into two scalar conduction problems.[20,21] Both $\boldsymbol{\sigma}_{eff}$ and the coupling tensor $\mathbf{L}_{eff}$ are then symmetric for every two-phase geometry. Under a three-fold or higher rotation axis, $\boldsymbol{\sigma}_{eff}$ is also isotropic in the plane perpendicular to the axis, so $\mathbf{S}_{eff} = \boldsymbol{\sigma}_{eff}^{-1}\mathbf{L}_{eff}$ stays symmetric and $S_a = 0$ exactly. A *p*4 composite containing only p-type and n-type phases with ideal contacts cannot produce $S_a$. A resistive interface, however, can supply a third, independent transport matrix: with contact resistivities of $10^{-9}$–$10^{-7}$ Ω m$^2$, a 1 mm two-phase p4 cell gives $|S_a|$ = 1.8–16.8 μV K$^{-1}$, reversed in its mirror image (Supplementary Table 1). A third phase removes this reduction when its matrix is linearly independent of the other two, and $\mathbf{L}_{eff}$ can then acquire an antisymmetric part (Methods, Supplementary Fig. 1, Supplementary Fig. 2). The condition is necessary, not sufficient: in two dimensions $\mathbf{L}_{eff}$ stays symmetric if all phases share one product σκ.[19] Milgrom identified third-order skew invariants;[23] here the cubic axial coefficient about one phase is proportional to $\Delta_{123}(\theta_1 G_{112} + \theta_2 G_{122})$: the phase-count determinant times $\theta_i$-weighted handed three-point correlations G (Methods).

Figure 2 tests these statements on a pinwheel cell of p-type, n-type and weakly conducting stripes. The two-phase cell gives $S_a = 0$ within $10^{-13}$ μV K$^{-1}$ (Fig. 2a,e). The three-phase cell with the same *p*4 symmetry gives $S_a$ between −15.65 and −15.43 μV K$^{-1}$ (Fig. 2b,e). Its mirror image gives the opposite sign (Fig. 2c), and a mirror-symmetric arrangement of the same three phases gives zero (Fig. 2d). In two-phase cells of lower symmetry, $\mathbf{S}_{eff}$ can still acquire a small antisymmetric part: the two decoupled scalar problems have different contrasts, so the symmetric tensors $\boldsymbol{\sigma}_{eff}$ and $\mathbf{L}_{eff}$ have misaligned principal axes (Supplementary Table 2).

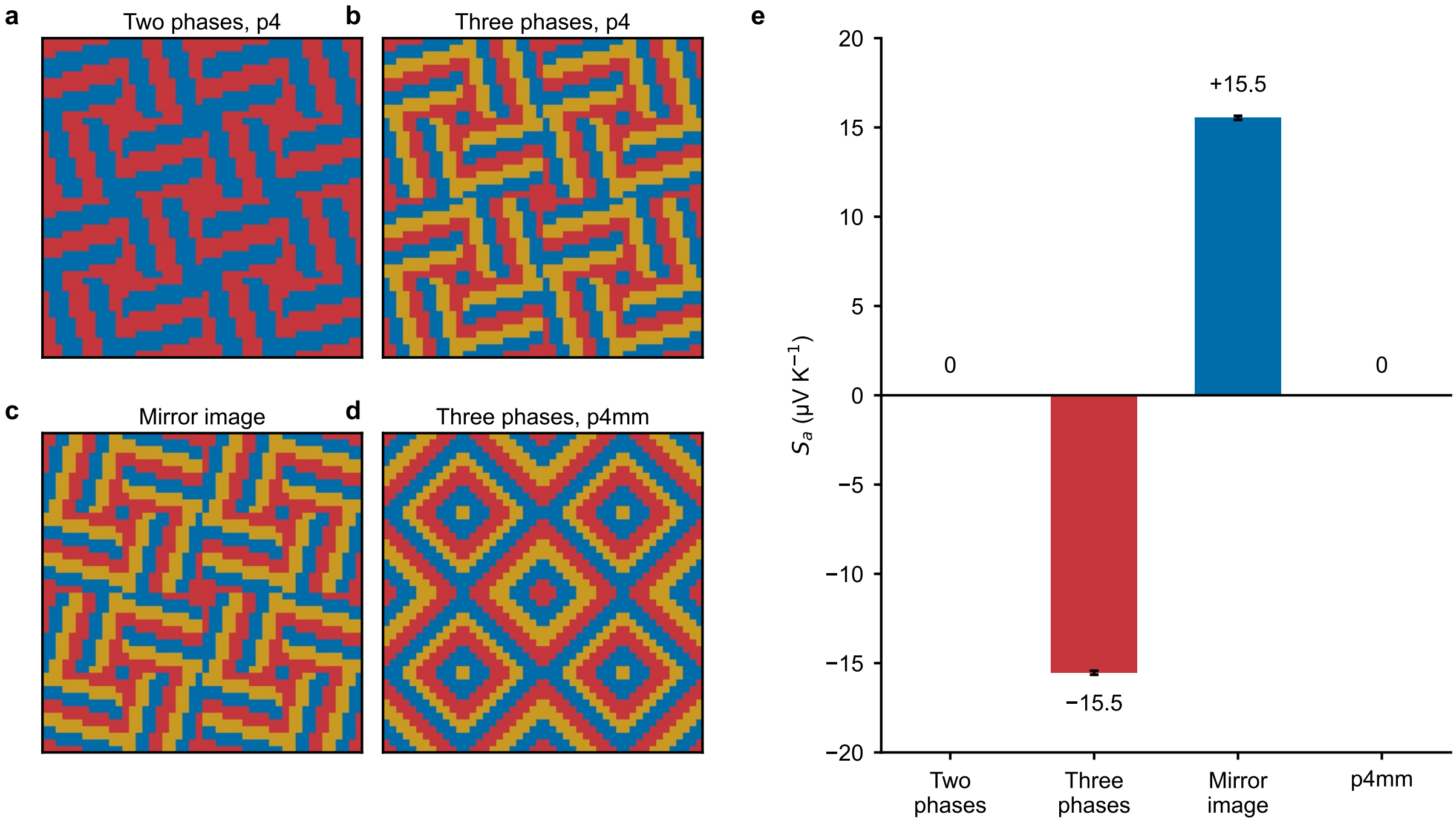

**Fig. 2 | Phase count and symmetry decide the axial thermoelectric response. a–d**, Two-by-two tiles of pinwheel unit cells. **a**, Two phases (p-type, red; n-type, blue) with p4 symmetry. **b**, Three phases with p4 symmetry; the third phase (ochre) conducts charge weakly. **c**, Mirror image of **b**. **d**, Three phases with p4mm symmetry. **e**, Axial Seebeck coefficient $S_a$ of the four cells. Bars give the corner-enriched upper-trial value and error bars the certified continuum interval; the zeros are finite-volume values below $10^{-13}$ μV K$^{-1}$. p and n phases: $\sigma = 10^5$ S m$^{-1}$, $S = \pm 200$ μV K$^{-1}$, $\kappa = 1.0$ W m$^{-1}$ K$^{-1}$; weak phase: $10^3$ S m$^{-1}$, 0 μV K$^{-1}$, 0.2 W m$^{-1}$ K$^{-1}$.

**Certified rather than extrapolated**

Pixelated cells contain corners where several phases meet, and the potential gradients there can be singular.[28] On the 24 × 24 pinwheel raster the energy error of nodal elements decays only as $h^{0.25}$,[29] and their $S_a$ remains outside the certified interval at 16-fold subdivision (Extended Data Fig. 1a). For $S_{xx}$, finite elements and finite volumes approach the continuum value from opposite sides, and Richardson extrapolation fails for both (Extended Data Fig. 1b).

We therefore bound the continuum values instead of extrapolating them. Two variational principles enclose the exact continuum effective matrix of the stated pixel geometry: any admissible potential field overestimates its energy, and any admissible flux field underestimates it. A corner-enriched conforming discretization gives the upper bound on the effective energy matrix. Extending planar conductivity duality to the coupled problem,[30] a variational upper bound for the dual problem gives a lower bound for the primal matrix after inversion and a 90° rotation.[22] For the pinwheel cell, the two bounds confine $S_a$ to [−15.65, −15.43] $\mu V\ K^{-1}$ and $S_{xx}$ to [−4.20, −3.97] $\mu V\ K^{-1}$ (Extended Data Fig. 1, Supplementary Table 3). The scheme reproduces the exact checkerboard conductivity at a contrast of $10^4$ to within 0.04% (Supplementary Note 4, Supplementary Fig. 3). Every interval quoted below is certified in this way or by the three-dimensional analogue (Methods); other values are point estimates.

**Charge and heat take different paths**

For weak thermoelectric coupling, $S_a$ reduces to the cell average of $\sigma S(\mathbf{E}_\sigma^{(x)}\cdot\mathbf{E}_\kappa^{(y)} - \mathbf{E}_\sigma^{(y)}\cdot\mathbf{E}_\kappa^{(x)})$, divided by $2\sigma_{eff}$ (Methods). Here $\mathbf{E}_\sigma^{(i)}$ and $\mathbf{E}_\kappa^{(i)}$ are the local fields of the uncoupled electrical and thermal conduction problems under a unit macroscopic gradient along i. The axial response therefore requires that charge and heat follow different paths, and that the Seebeck coefficient varies along this difference with a fixed handedness. When $\kappa/\sigma$ is uniform, the two paths coincide and this leading term vanishes, leaving only contributions of higher order in S.

An exact discrete decomposition shows where the response is generated in an optimized design (design A, Fig. 3a). Its third phase conducts heat nearly as well as the thermoelectric phases but charge $10^4$ times worse, so charge detours around it while heat crosses it (Fig. 3b). In this cell, on a 128 × 128 raster, 86% of the axial integral comes from the interior of the n-type arms (Fig. 3c). A third phase that passes heat but blocks charge is therefore the central design rule. It can be an electrical insulator: with $\sigma = 10^{-6}$ S m$^{-1}$ and $\kappa = 0.3$ or 1.0 W m$^{-1}$ K$^{-1}$, design A gives certified $S_a$ of 44.0–44.6 and 53.1–53.8 $\mu V\ K^{-1}$.

**Tuning sign and magnitude**

Rotating-square cells make the role of handedness explicit: with p-type and n-type squares rotating oppositely in a soft conducting matrix, $S_a$ is odd in the rotation angle and vanishes where mirror symmetry is restored (Fig. 3d), so rotating-square mechanisms[31,32] could switch its sign. The matrix conductivity sets the sign as well (Fig. 3e).

Optimization over geometry and materials raises the response substantially (Fig. 3f, Supplementary Table 3). Design A, a three-colour pinwheel tiling of p-type, n-type and thermally conducting phases, gives $S_a$ = 49.5–50.2 μV $K^{-1}$ on its 64 × 64 raster; a 200 × 200 raster of the same drawing, bounded less tightly, gives 43–55 μV $K^{-1}$ (Supplementary Table 4). On the 64 × 64 raster this is 26% of the Seebeck half-span of its constituents, with $\sigma_{eff}$ = 3.2 × $10^4$ S $m^{-1}$. Its transverse power factor $S_a^2\sigma_{eff}$ reaches 78–82 μW $m^{-1}$ $K^{-2}$, and with $\kappa_{eff}$ = 1.24 W $m^{-1}$ $K^{-1}$ its transverse figure of merit is $zT_a$ = 0.019. The thermally conducting third phase that creates the response thus also limits conversion efficiency, so the effect suits sensing and junction-free circulation rather than power generation. A three-phase design with a carbon phase reaches 29%, and adding 4.7% copper raises this to 32%, or $S_a$ = −61.5 to −61.1 μV $K^{-1}$. This is about ten times the anomalous Nernst coefficient of iron-based magnets, about 6 μV $K^{-1}$ at room temperature,[2] as a signal per temperature gradient rather than as conversion efficiency.

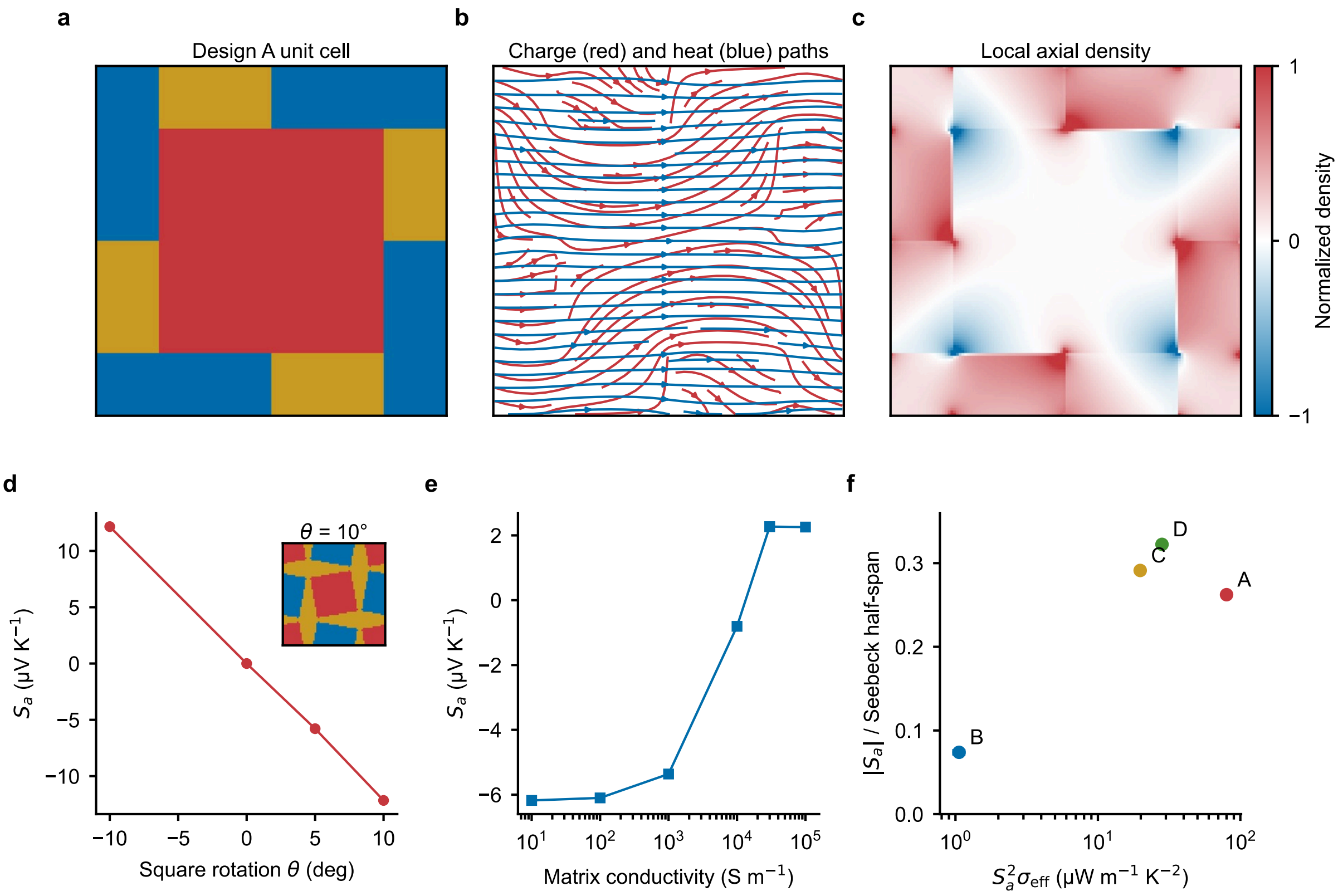

**Fig. 3 | Mechanism and design space. a**, Unit cell of design A: p-type core (red), n-type arms (blue) and a thermally conducting, electrically weak phase (ochre). **b**, Streamlines of the electrical current (red) and the conductive heat flux (blue) of the uncoupled conduction problems under a unit macroscopic gradient along x. **c**, Local density of the exact face-wise decomposition of the axial coefficient, normalized by the 99th percentile of its magnitude; its cell average equals $\sigma_{eff}S_a$. Finite-volume calculation on a 128 × 128 raster. **d**, $S_a$ of rotating-square cells (inset, θ = 10°) versus the square rotation angle θ for $|\theta| \leq 10°$, below the angle at which neighbouring squares nearly touch; square side 0.8 of the half-cell, carbon-composite matrix. **e**, $S_a$ versus the conductivity of the matrix phase ($S = 10$ μV K$^{-1}$, $\kappa = 0.3$ W m$^{-1}$ K$^{-1}$) at θ = 16° and square side 0.7. Panels **d** and **e** are finite-volume point estimates. **f**, Certified designs: $|S_a|$ normalized by the Seebeck half-span $(\max S_i - \min S_i)/2$ of the constituents versus the transverse power factor $S_a^2\sigma_{eff}$. A, three-colour pinwheel tiling; B, rotating squares; C, three-phase design with carbon; D, four-phase design with copper. Intervals are floating-point variational enclosures for the stated rasters, not statistical error bars.

### A line heat source acts like a current-carrying wire

Without circulating current, a radial heat flow q′ per unit length around a line source creates the gradient $\nabla T = -q'/(2\pi\kappa_{eff}r)\,\hat{r}$. The axial response converts this into an azimuthal field $E_\theta = S_a q'/(2\pi\kappa_{eff}r)$, which has the form of the magnetic field of a straight wire. The electromotive force around any closed path that winds n times around the source is therefore $n\cdot S_a q'/\kappa_{eff}$: the driving field is curl-free away from the source but not conservative around it. We tested this law on spiral strips carrying the conforming-bound effective tensor of design A, embedded in an electrically insulating matrix of equal thermal conductivity. After the longitudinal term $S_{xx}\Delta T$ between the two ends is subtracted, the open-circuit voltage is 205, 410 and 615 μV for n = 1, 2 and 3 at a temperature difference of 1 K between the line source (radius 2 mm) and the outer boundary (radius 9.3 mm) (Fig. 4b,d). Reversing the winding or the handedness reverses the sign. The driving field $\mathbf{S}_{\text{eff}}\nabla T$ has curl $-S_a(\nabla^2 T)\hat{z}$, which vanishes where the temperature is harmonic, as in the current-free strip; the matrix still changes the thermal boundary conditions. At a matrix/strip thermal-conductivity ratio of 0.03, the law holds within 1% on the finest grid with q′ the source heat input (Supplementary Table 5); at ratios 0.003 and 10, deviations are 5–12%, while $S_a$ times the line integral of the rotated temperature gradient predicts the corrected voltage within 1.1% (Supplementary Note 8).

In a closed annulus, the same transverse electromotive force drives a circulating current without any junction. For an annulus of the pinwheel cell of Fig. 2b with 4 and 9 mm radii and 1 mm thickness, the microstructured model gives $2.8 \times 10^{-4}$ A per kelvin of radial temperature difference. With 12 or more cells across the box it agrees within 11% with a homogeneous annulus carrying the effective tensor of the same mesh (Fig. 4a,c); with the certified tensor the homogeneous annulus carries $2.56–2.61 \times 10^{-4}$ A $K^{-1}$. The certified current produces 28 nT $K^{-1}$ at the centre, far above the picotesla sensitivity of SQUID magnetometers used to detect thermoelectric currents,[33] and the mirror-image annulus reverses it. Inhomogeneous thermoelectrics can carry local eddy currents,[34,35] but a mirror-symmetric annulus of the same three phases carries a net current below $10^{-13}$ A. Spiral thermoelements sum the eddy voltage of a layered plate over loops;[36] here it follows from a periodic effective tensor.

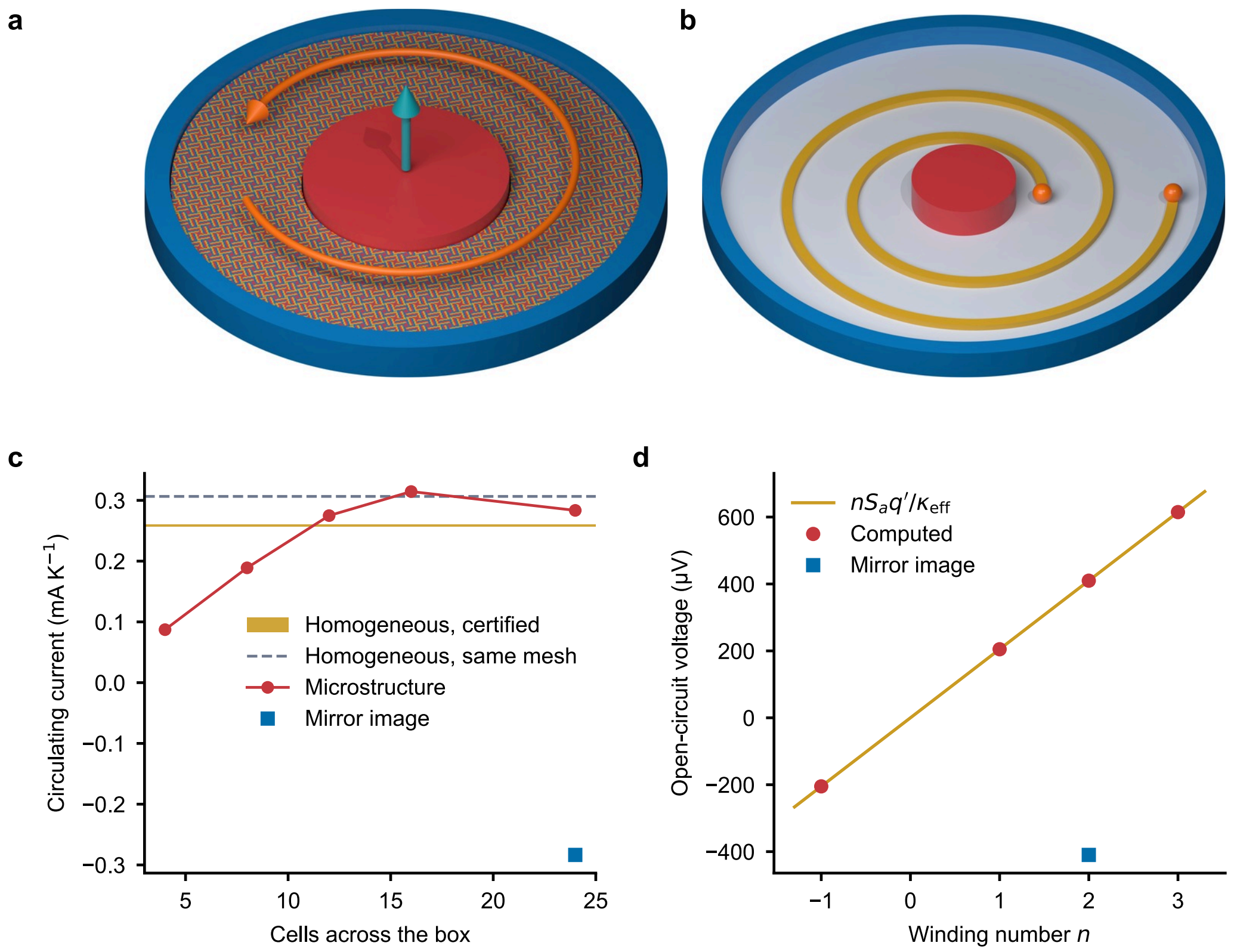

**Fig. 4 | Circulating currents and winding-number voltages. a**, Annulus of the pinwheel cell of Fig. 2b (radii 4 and 9 mm, 1 mm thick, 24 cells across the 20 mm box) around a hot core inside a cold ring, with the circulating current I (orange) and the central field $B_z$ (teal). **b**, Spiral strip winding twice around a line heat source (radius 2 mm) inside a cold boundary (radius 9.3 mm); the terminals are orange. Panels **a** and **b** are renderings of the computed geometries. **c**, Circulating current per kelvin in the annulus versus the number of cells across the simulation box. The narrow ochre band is the homogeneous prediction with the certified effective tensor, and the dashed line is the homogeneous annulus on the same mesh with the effective tensor of the same 24 × 24 bilinear cell discretization, which overestimates $\sigma_{eff}$ by 19%; the blue square is the mirror-image annulus. **d**, Longitudinally corrected open-circuit voltage of spiral strips that wind n times around the line source at 1 K temperature difference. The ochre line is $nS_aq'/\kappa_{eff}$; the blue square is the mirror image at n = 2. Panels **c** and **d** are finite-element point estimates.

**Colour symmetry selects printable architectures**

In three dimensions the axial vector requires a pyroaxial point group such as 4, 4/m (mirror perpendicular to the axis) or 6. The relevant symmetry is the colour group of the arrangement, the operations that leave every phase in place, rather than that of the bare geometry. Straight-rod woodpiles, native to extrusion printing, illustrate the distinction: every one we generated kept two-fold axes perpendicular to the stacking direction (colour group 422), and its axial coefficient vanished (Fig. 5c). Dividing each rod into p-type, n-type and weak segments of unequal length removes these axes: of 82 segmented woodpiles, 46 have colour group 4 and 35 the group m with a mirror perpendicular to the stacking axis, both of which allow the axial vector along z, and one retains 422.

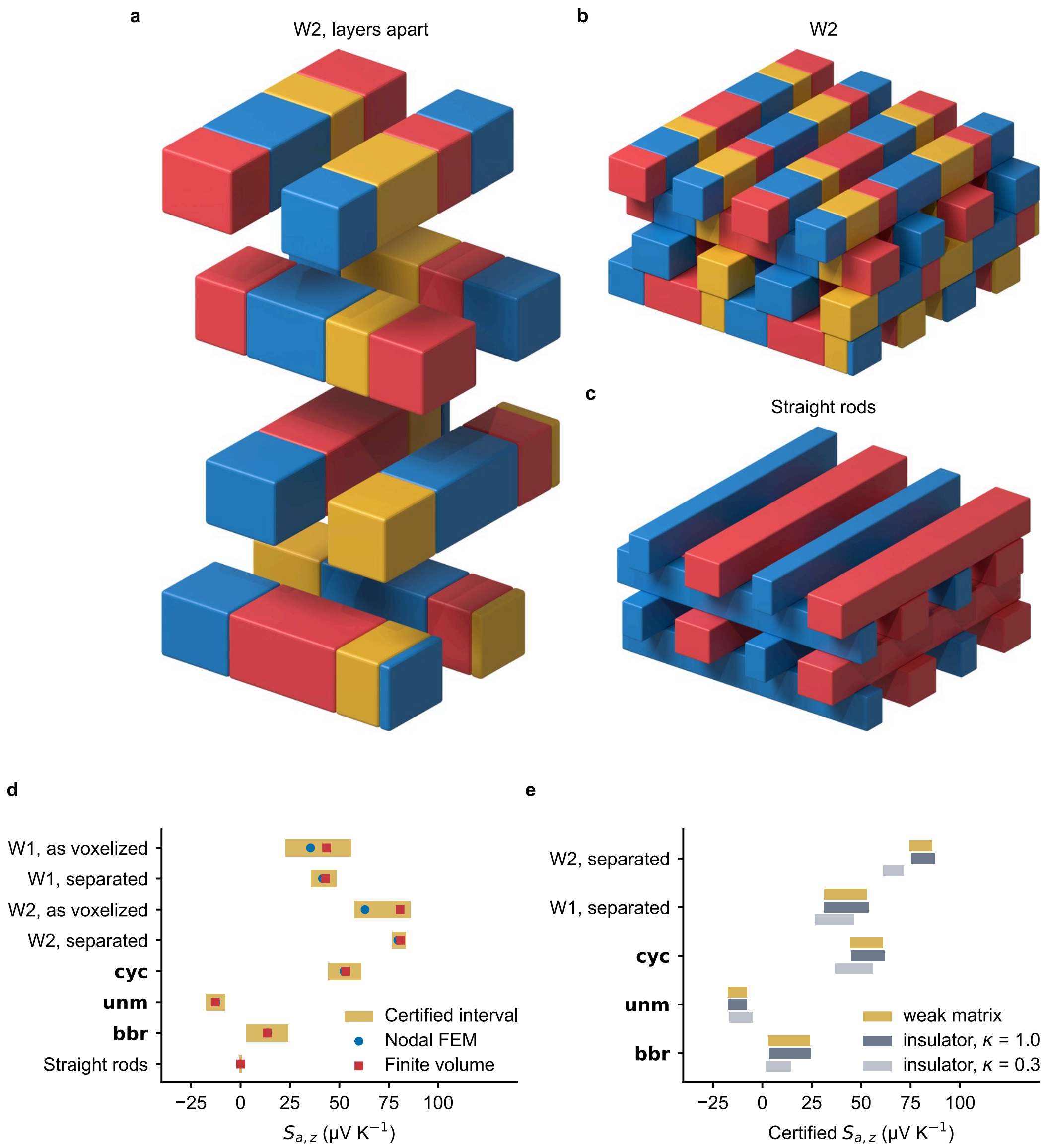

a
W2, layers apart
b
W2
c
Straight rods
d
W1, as voxelized
W1, separated
W2, as voxelized
W2, separated
cyc
unm
bbr
Straight rods
Certified interval
Nodal FEM
Finite volume
−25
0
25
50
75
100
$S_{a,z}$ (μV K$^{-1}$)
e
W2, separated
W1, separated
cyc
unm
bbr
weak matrix
insulator, κ = 1.0
insulator, κ = 0.3
−25
0
25
50
75
100
Certified $S_{a,z}$ (μV K$^{-1}$)

**Fig. 5 | Printable woodpiles. a**, Segmented-rod woodpile W2, one cell with its four layers drawn apart. **b**, W2 as a 2 × 2 block (colour group 4; $S_{a,z}$ = 76–84 μV $K^{-1}$ with separated junctions, **d**). **c**, Straight-rod woodpile as a 2 × 2 block (colour group 422, $S_{a,z} = 0$). p-type red, n-type blue, weak rod segments ochre; the weak matrix, the same material as the ochre segments, fills the space between the rods and is not drawn. The renderings use the design parameters, which reproduce the computed 24-voxel label maps exactly, and show W2 before junction separation, which changes 0.55% of the volume (Extended Data Fig. 3). **d**, Certified continuum intervals (ochre) for fixed voxel geometries, floating-point variational enclosures at the finest subdivision; for W2 with separated junctions at twofold subdivision, ball arithmetic bounding every rounding error gives 74–86 μV $K^{-1}$ (Methods), with nodal finite-element (blue) and finite-volume (red) point estimates at the finest subdivision. W1 is the atlas woodpile (16-voxel cell) from which W2 was optimized; W2 itself is not one of the 82 atlas woodpiles. **cyc**, **unm** and **bbr** are voxel structures derived from the corresponding RCSR nets with their original c/a ratios; signed values are shown for the configurations computed, and mirror images reverse every sign. **e**, Certified intervals with the weak matrix (ochre) replaced by an electrical insulator ($\sigma = 10^{-6}$ S $m^{-1}$) of thermal conductivity 1.0 (dark grey) or 0.3 W $m^{-1}$ $K^{-1}$ (light grey); voxels refined twofold, fourfold for **bbr** at 0.3 W $m^{-1}$ $K^{-1}$.

The finite-volume estimates obeyed the colour-symmetry selection rule in all 215 structures tested: 82 segmented and 17 straight woodpiles and 116 tetragonal nets of the Reticular Chemistry Structure Resource (Fig. 6, Supplementary Table 6).[25] Net struts were coloured by their in-plane orientation modulo 90°, a handed rule that breaks mirrors containing the axis and perpendicular two-fold axes whenever struts lie off the 0° and 45° directions. Allowed structures gave $|S_{a,z}|$ from 0.06 to 53 μV $K^{-1}$, the smallest values unresolved; forbidden ones stayed below 2 × $10^{-8}$ μV $K^{-1}$ (Extended Data Fig. 2). Phase count alone does not decide: recoloured by the folded strut angle, which restores the parent symmetry, 24 nets that the handed rule made axial stay three-phase but become forbidden, and all 24, including **unm** and **bbr**, give $|S_{a,z}|$ below $10^{-10}$ μV $K^{-1}$ (Supplementary Table 7, Extended Data Fig. 4).

Certified intervals fix the magnitude of selected structures (Fig. 5d, Supplementary Table 8). The optimized segmented woodpile W2 (Fig. 5a,b) as voxelized gives $S_{a,z}$ between 57 and 87 μV $K^{-1}$, and its mirror image reverses the sign. There like phases meet partly along voxel edges, where nodal finite elements and finite volumes still differ by 22% at the finest subdivision; separating these contacts by weak-phase gaps two voxels wide changes 0.55% of the volume (Extended Data Fig. 3). The certified interval then narrows to 76–84 μV $K^{-1}$, with $\sigma_{eff,xx}$ = 5.3–5.6 × $10^3$ S $m^{-1}$ and a transverse power factor of 31–39 μW $m^{-1}$ $K^{-2}$, and the two methods agree within 1.7% (Fig. 5d, Supplementary Table 9). Evaluated with ball arithmetic, which bounds every rounding error, the same bounds give 74–86 μV $K^{-1}$ on the twofold-refined voxels (Methods). The structures derived from the nets **cyc**, **unm** and **bbr**, voxelized with their original c/a ratios, give certified $S_{a,z}$ of 44 to 62, −18 to −7 and 2 to 25 μV $K^{-1}$, with struts filling only 13–23% of the cell. All five structures keep p–n face connectivity in three directions and non-zero certified $S_{a,z}$ when the weak matrix is replaced by an electrical insulator; for W2 with κ = 1.0 W $m^{-1}$ $K^{-1}$ it gives 75–88 μV $K^{-1}$ (Fig. 5e, Supplementary Table 10).

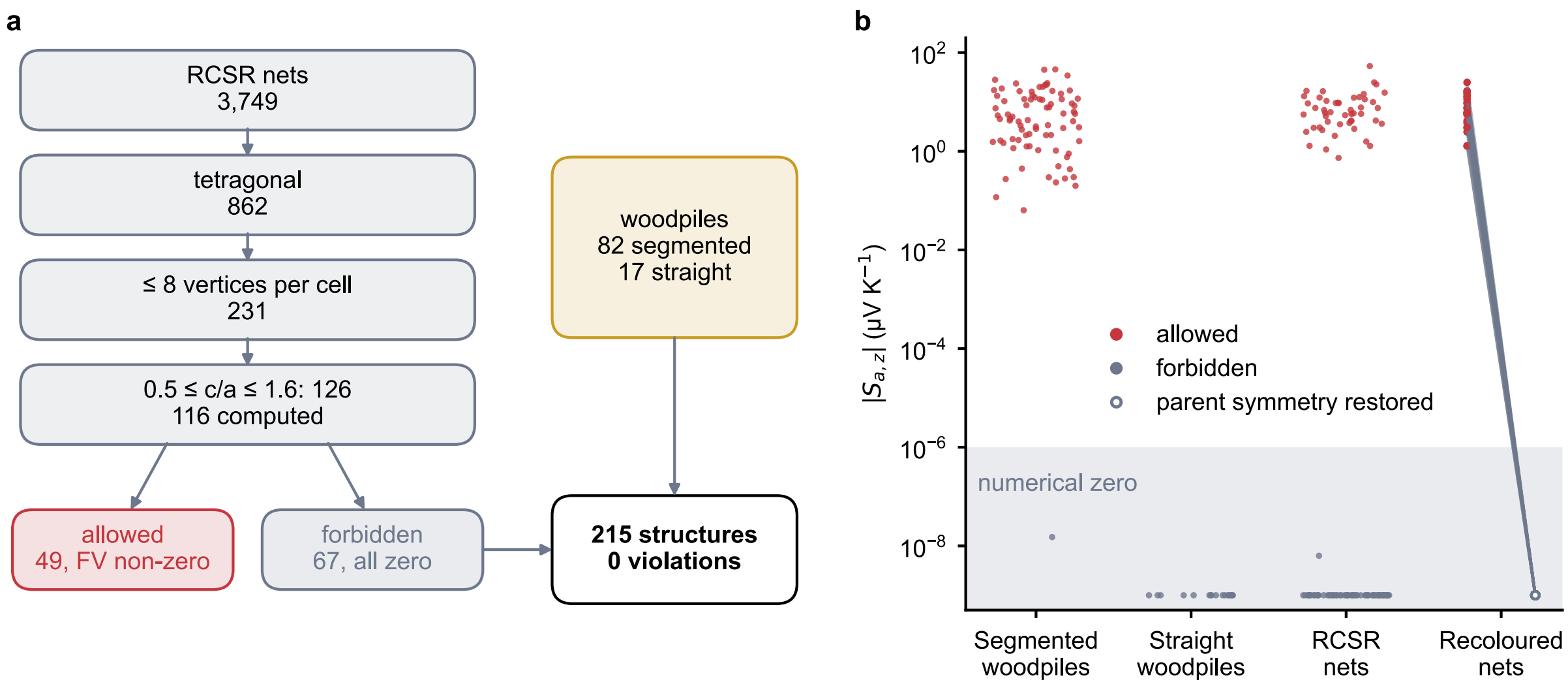

**Fig. 6 | Colour symmetry constrains the axial response across crystal nets. a**, Selection of nets from the Reticular Chemistry Structure Resource and the colour-group verdicts of the 116 computed nets: the 49 allowed nets give $|S_{a,z}|$ = 0.73–53.3 μV $K^{-1}$ and the 67 forbidden nets, all of them single-coloured and therefore two-phase, give $|S_{a,z}| < 10^{-8}$ μV $K^{-1}$ (all nets in Supplementary Table 16). Struts are coloured p-type for in-plane orientations of 0–45° modulo 90° and n-type for 45–90°; vertical struts and struts at exactly 0° or 45° are p-type, so nets with only such struts are single-coloured and keep their mirrors. **b**, $|S_{a,z}|$ of all 215 structures, grouped by family and coloured by whether their colour group allows (red) or forbids (grey) it. Finite-volume (FV) values on cubic voxels, 16 or 24 along a, so that net-derived cells have c/a rounded to a multiple of 0.25; forbidden cases fall at numerical zero. The 49 allowed structures are drawn in Extended Data Fig. 2. The column of recoloured nets pairs the 24 nets that the folded-angle rule makes forbidden while keeping three phases: handed colouring (red) and parent-symmetry colouring (open circles, plotted at $10^{-9}$ μV $K^{-1}$; computed residuals below $10^{-10}$); phase fractions differ between the two colourings (Supplementary Table 7).

## Design rules and practical constraints

An antisymmetric thermopower does not require magnetism. In composites of isotropic phases under a three-fold or higher axis, two phases with ideal interfaces cancel exactly, and a response requires a ferroaxial-type arrangement and three linearly independent transport matrices. The symmetry allowance,[16] the non-symmetric homogenized Seebeck tensor, including an axial term for polycrystals of anisotropic crystallites,[19] the two-phase reduction[20,21] and the skew invariants of coupled fields[23] were known, but none gave a composite of isotropic phases with isotropic in-plane conduction and an axial thermopower. New here are the constructive escape from that reduction under isotropic in-plane conduction, the factorization of the leading skew term into the phase-count determinant and material-weighted handed correlations, phase-preserving colour symmetry as a test for an antisymmetric effective tensor with crystal nets as a design space, and certified magnitudes. Nothing in the phase-count argument is specific to thermoelectricity: the obstruction, its three-phase escape and the weak-contrast onset hold for any reciprocal two-current composite with isotropic, positive-definite local matrices and ideal interfaces,[21,23] so thermodiffusion and closed Darcy-scale electrokinetics can carry the same axial cross response (Supplementary Note 10).

Design A keeps $S_a$ above 46 μV $K^{-1}$ when its phase junctions are separated by 1–4% of the cell or its corners are rounded by up to 5%, and contact resistivity up to $10^{-6}$ Ω $m^2$ changes $S_a$ by less than 20% (Supplementary Table 4, Supplementary Table 11). The defining measurement is the half-difference of $S_{xy}$ and $S_{yx}$ taken with orthogonal gradients on one specimen with matched probes and terminations: for design A with separated junctions and centred probes it recovers 96.0–99.5% of the certified bulk value on N × N specimens with N = 4–16, and 74.6–99.5% with misplaced probes; the pinwheel of Fig. 2b gives 71.9–133.4% (Supplementary Table 12, Supplementary Table 13). Reflection in a plane containing the axis reverses the bulk coefficient, so mirror-image specimens with mirrored probes and terminations, and recoloured lattices, serve as null tests; finite two-phase rings, by contrast, can carry boundary currents larger than the signal (Supplementary Note 6). For a 10 K drive, linear-response estimates give about 0.5 mV across a square of design A, transverse to a uniform gradient, 6 mV of longitudinally corrected voltage along its three-turn spiral in a matched matrix and 0.3 μT at the centre of the pinwheel annulus; the mirror-image material at fixed device geometry reverses each signal. Extrusion printing of chalcogenide inks already reaches this scale,[26,27] but three materials per cell require multimaterial patterning. Sensors that read a heat flow through a winding number, and junction-free circulating currents, are possible applications.

## Methods

**Transport model.** Each phase is isotropic with constant σ, S and κ at $T_0$ = 300 K; Joule and Thomson heating are neglected in linear response. The fluxes are related to the forces by a local matrix that is symmetric by Onsager reciprocity and positive definite,[15,22]

$$(\mathbf{j}, \mathbf{q}/T_0) = (s \otimes I)(-\nabla V, -\nabla T), \quad s = [\sigma, \sigma S; \sigma S, (\kappa + T_0\sigma S^2)/T_0], \tag{1}$$

where $s \otimes I$ acts identically on every spatial component. Effective coefficients are defined from volume-averaged fluxes,[22] as in asymptotic homogenization of layered thermoelectric composites.[37] Periodic homogenization then gives the effective energy matrix, 4 × 4 in two dimensions and 6 × 6 in three, with blocks $\boldsymbol{\sigma}_{eff}$, $\mathbf{L}_{eff}$, $\mathbf{L}_{eff}^T$ and the thermal block $\mathbf{M}_{eff}$, from which the Seebeck tensor and the open-circuit thermal conductivity follow as

$$C = [\boldsymbol{\sigma}_{eff}, \mathbf{L}_{eff}; \mathbf{L}_{eff}^T, \mathbf{M}_{eff}], \quad \mathbf{S}_{eff} = \boldsymbol{\sigma}_{eff}^{-1}\mathbf{L}_{eff}, \quad \boldsymbol{\kappa}_{eff} = T_0(\mathbf{M}_{eff} - \mathbf{L}_{eff}^T\boldsymbol{\sigma}_{eff}^{-1}\mathbf{L}_{eff}), \tag{2}$$

with $S_a = (S_{xy} - S_{yx})/2$ (Supplementary Note 1). Reciprocity makes the whole matrix symmetric but leaves the spatial symmetry of the off-diagonal block $\mathbf{L}_{eff}$ free; in single-field conduction, by contrast, the symmetric effective conductivity admits no antisymmetric part. For two phases, a uniform congruence transformation of (V, T) diagonalizes $s_1$ and $s_2$ simultaneously.[20–22] Three or more phase matrices cannot in general be diagonalized simultaneously; they can when all are linear combinations of two of them,[22] and the two-phase isomorphism does not extend to three phases.[38] Because the symmetric 2 × 2 matrices form a three-dimensional space, the converse also holds. For three phases the reduction is avoided exactly when $s_1$, $s_2$ and $s_3$ span the three-dimensional space of symmetric 2 × 2 matrices, that is, when the phase-count determinant, with one row per phase, is non-zero:

$$\Delta_{123} = \det[\sigma_i, \sigma_i S_i, \kappa_i + T_0\sigma_i S_i^2]_{i = 1, 2, 3} \neq 0. \tag{3}$$

This condition is necessary and sufficient to prevent simultaneous diagonalization by congruence, but not sufficient for a spatially antisymmetric coupling block. In two dimensions, phases with equal $\det s_i = \sigma_i\kappa_i/T_0$ lie, after one common normalization, on the exact relation {Λ $\otimes$ P : det Λ det P = 1},[19] so $C = \Lambda^* \otimes P^*$ for every geometry and $\mathbf{S}_{\mathrm{eff}} = (\Lambda^*_{12}/\Lambda^*_{11})I$ is scalar even when $\Delta_{123} \neq 0$ (Supplementary Note 3). Onsager reciprocity of the effective matrix holds to better than $10^{-11}$ in all calculations.

**Finite-volume and finite-element solvers.** Cell-centred finite volumes used the matrix harmonic mean $2b(a + b)^{-1}a$ on faces between phases a and b, the face matrix that is exact for a laminate. Nodal bilinear finite elements served as an independent method. Both reproduced laminate solutions to $10^{-14}$.

**Certified bounds.** Corner singularities were resolved by circular subdomains at every multi-phase vertex, inside which the two coupled potentials are expanded in angular eigenmodes with radial behaviour $r^{\lambda}$, following the singular-function enrichment of finite elements[39] and the scalar theory of multimaterial corners[28] (Supplementary Note 4). This $H^1$-conforming scheme bounds the effective energy matrix from above. The two-dimensional dual problem, solved with the inverse material matrix, gives an upper bound $U_{\mathrm{dual}}$ on the dual matrix and hence the lower bound

$$L = (I_2 \otimes J)^T U_{\mathrm{dual}}^{-1}(I_2 \otimes J), \quad J = [0, 1; -1, 0], \tag{4}$$

where J is the 90° rotation in the plane; this extends planar duality[30,40] and the complementary variational principle[22,41] to the coupled periodic cell. Two-sided variational bounds of this kind are established for scalar periodic homogenization;[42] here they are applied to the coupled energy matrix. With the upper and lower bounds U and L on the effective energy matrix C in the Loewner order, $L \preceq C \preceq U$, intervals for $S_a$, $S_{xx}$, $\sigma_{\mathrm{eff}}$ and derived quantities follow from optimization over all symmetric matrices in this set that obey the verified four-fold symmetry. Writing $C = L + D^{1/2}XD^{1/2}$ with $D = U - L$ and $0 \preceq X \preceq I$, the support function of this set is

$$F(H) = \max_{L \preceq C \preceq U} \mathrm{tr}(HC) = \mathrm{tr}(HL) + \Sigma_{\lambda > 0}\, \lambda(D^{1/2}HD^{1/2}), \tag{5}$$

the second term being the sum of the positive eigenvalues. Under four-fold symmetry, for the ordering $(V_x, V_y, T_x, T_y)$,

$$\boldsymbol{\sigma}_{eff} = aI,\ \ \mathbf{L}_{eff} = bI + cJ,\ \ a = (C_{11} + C_{22})/2,\ \ b = (C_{13} + C_{24})/2,\ \ c = (C_{14} - C_{23})/2, \qquad (6)$$

so that $S_{xx} = b/a$ and $S_a = c/a$. Because L, U and these functionals are invariant under four-fold rotation, the extrema over the Loewner set are attained on its symmetric part, and because ratios of linear functionals with a positive denominator are quasilinear,[43] the upper and lower limits are the roots t of $F(H_{b,c} - tH_a) = 0$ and $F(tH_a - H_{b,c}) = 0$. The scheme reproduces the exact checkerboard conductivity $(\sigma_1\sigma_2)^{1/2}$, which follows from the phase-interchange theorem,[30] at a contrast of $10^4$ to within 0.04%. The enclosures are variational bounds for the stated pixel geometry; quadrature and rounding errors are bounded only where ball arithmetic is stated.

**Weak-coupling expression and local decomposition.** Following the field-weighted averages for the thermopower of weakly coupled inhomogeneous media,[44] expanding the tensor problem to first order in S gives $L_{eff,ij} \approx \langle \sigma S\, \mathbf{E}_\sigma^{(i)}\cdot \mathbf{E}_\kappa^{(j)} \rangle$ and hence, for a cell with isotropic $\boldsymbol{\sigma}_{eff}$,

$$S_a \approx \langle \sigma S(\mathbf{E}_\sigma^{(x)}\cdot \mathbf{E}_\kappa^{(y)} - \mathbf{E}_\sigma^{(y)}\cdot \mathbf{E}_\kappa^{(x)})\rangle/(2\sigma_{eff}). \qquad (7)$$

For finite coupling, the cross block was written as $C^{VT}_{ij} = \Sigma_f\, \Delta\mathbf{\Phi}_f^{(V,i)}\cdot s_f \Delta\mathbf{\Phi}_f^{(T,j)}/|\text{cell}|$, a sum over faces f of the solved cell problems. Here $\Delta\Phi_f^{(\alpha,i)}$ is the jump of the coupled potentials (V, T) across face f in the cell problem driven by a unit gradient of field α along i, and the axial density is the face-wise value of $(C^{VT}_{xy} - C^{VT}_{yx})/2$. The decomposition reproduces the flux-based value to within $10^{-12}$.

**Weak-contrast expansion.** Let $s_1$, $s_2$ and $s_3$ denote fixed endpoint matrices. We introduce $s(\mathbf{x}; \varepsilon) = s_3 + \varepsilon[\chi_1(\mathbf{x})\Delta_1 + \chi_2(\mathbf{x})\Delta_2]$, where $\Delta_i = s_i - s_3$ and $\chi_i$ is the indicator of phase i. The reference is $K_0 = s_3 \otimes I$, and $\delta K = \delta s \otimes I$ with $\delta s = s - s_3$. Periodic homogenization gives the perturbation series[22,23] (derivation in Supplementary Note 7)

$$C = K_0 + \langle \delta K\rangle - \langle \delta K\, \Gamma\, \delta K\rangle + \langle \delta K\, \Gamma\, \delta K\, \Gamma\, \delta K\rangle - \ldots,\quad \hat{\Gamma}(\mathbf{k}) = s_3^{-1} \otimes \hat{\mathbf{k}}\hat{\mathbf{k}}^T, \qquad (8)$$

with $\hat{\Gamma}(0) = 0$. Angle brackets denote cell averages, and $\mathbf{k}$ runs over the reciprocal lattice. Fourier coefficients are normalized by the cell volume. The first-order term is spatially isotropic. The second-order term, $-\Sigma_{\mathbf{k} \neq 0}\, \delta\hat{s}(-\mathbf{k})s_3^{-1}\delta\hat{s}(\mathbf{k}) \otimes \hat{\mathbf{k}}\hat{\mathbf{k}}^T$, contains only symmetric spatial projectors. Neither term contributes to $c = (C_{Vx,Ty} - C_{Vy,Tx})/2$. We use $J = [0, 1; -1, 0]$, fixing the sign of the axial coefficient. The third-order spatial factor is $(\hat{\mathbf{k}}\cdot\hat{\mathbf{q}})\hat{\mathbf{k}}\hat{\mathbf{q}}^T$; its antisymmetric part is $\frac{1}{2}(\hat{\mathbf{k}}\cdot\hat{\mathbf{q}})(\hat{\mathbf{k}}\times\hat{\mathbf{q}})\cdot\hat{z}\, J$. Relabelling $(\mathbf{k}, \mathbf{q})$ as $(-\mathbf{q}, -\mathbf{k})$ reverses this kernel and transposes the material product. Consequently, only the antisymmetric material product contributes to c; this is the third-order skew invariant of coupled-field composites with three or more phases.[23] Defining $c(\varepsilon) = \varepsilon^3 c_3 + O(\varepsilon^4)$, we obtain

$$c_3 = \tfrac{1}{2} \Sigma_{i,j,\ell=1}^{2} [\Delta_i R \Delta_j R \Delta_\ell]_{VT}\, G_{ij\ell}, \quad R = s_3^{-1}, \tag{9}$$

$$G_{ij\ell} = \mathrm{Re}\, \Sigma_{\mathbf{k},\,\mathbf{q}\neq 0}\, \hat{\chi}_i(-\mathbf{k})\, \hat{\chi}_j(\mathbf{k}-\mathbf{q})\, \hat{\chi}_\ell(\mathbf{q})\, (\hat{\mathbf{k}}\cdot\hat{\mathbf{q}})(\hat{\mathbf{k}}\times\hat{\mathbf{q}})\cdot\hat{z}. \tag{10}$$

The tensor satisfies $G_{ij\ell} = -G_{\ell ji}$. For pixelated cells, the sum is interpreted through a common symmetric Fourier cutoff. In two dimensions, $G_{ij\ell}$ changes sign under reflection and vanishes when the reflection preserves every phase label. For a general homogeneous reference $s_0$, the cubic contraction includes every phase contrast and uses $R = s_0^{-1}$. $\theta_i$, the trace part of the normalized contrast $s_3^{-1/2}\Delta_i s_3^{-1/2}$, is defined in equation (11). With $\Delta_{123}$ of equation (3), the determinant formed from the entries $(s_{i,11}, s_{i,12}, s_{i,22})$ is $\Delta_{123}/T_0$. For cells with a three-fold or higher rotation axis, $S_a = c/a$ with $a(0) = s_{3,11}$, giving

$$S_a = \varepsilon^3\, [\Delta_{123}/(T_0 s_{3,11} \det s_3)](\theta_1 G_{112} + \theta_2 G_{122}) + O(\varepsilon^4), \quad \theta_i = \tfrac{1}{2}\,\mathrm{tr}(s_3^{-1}s_i) - 1. \tag{11}$$

The cubic coefficient therefore contains a material determinant multiplying a material-weighted combination of two geometric correlations. For two phases, choosing one as reference leaves only $G_{111}$, which vanishes by endpoint antisymmetry, so $c = 0$; without a three-fold axis $S_a$ can still be non-zero (Supplementary Note 2). Linear dependence of the endpoint matrices gives $\Delta_{123} = 0$ and removes c, and hence $S_a$, throughout this phase-3 interpolation. If $\theta_1 G_{112} + \theta_2 G_{122} = 0$, the cubic term vanishes and $S_a = O(\varepsilon^4)$; the fourth-order coefficient may also vanish. Traceless normalized contrasts, $\theta_1 = \theta_2 = 0$, provide one such cancellation. In three dimensions, the geometric tensor becomes an axial vector by retaining $\hat{\mathbf{k}}\times\hat{\mathbf{q}}$ instead of its z component. Its z component determines $c_{z,3}$, and equation (11) applies to $S_{a,z}$ for cells with a three-fold or higher rotation axis along z. For the pinwheel of Fig. 2b, phase 1 is p-type, phase 2 is n-type and phase 3 is the weakly conducting phase. The closed form agrees with complete solutions of the cell problem in two and three dimensions, and at full contrast higher orders dominate (Supplementary Note 7 and Supplementary Fig. 4).

**Materials and design search.** At 300 K we used p-type BiSbTe ($\sigma = 10^5$ S m$^{-1}$, $S = 200$ μV K$^{-1}$, $\kappa = 1.2$ W m$^{-1}$ K$^{-1}$) and n-type $Bi_2Te_{2.7}Se_{0.3}$ ($10^5$ S m$^{-1}$, −180 μV K$^{-1}$, 1.3 W m$^{-1}$ K$^{-1}$). The weak phase is a carbon composite ($10^3$ S m$^{-1}$, 10 μV K$^{-1}$, 0.3 W m$^{-1}$ K$^{-1}$); Cu ($6 \times 10^7$ S m$^{-1}$, 1.8 μV K$^{-1}$, 400 W m$^{-1}$ K$^{-1}$) and Ni ($1.4 \times 10^7$ S m$^{-1}$, −19 μV K$^{-1}$, 90 W m$^{-1}$ K$^{-1}$) were also included. The third phase of design A, $\sigma = 10.45$ S m$^{-1}$, $S = 10$ μV K$^{-1}$ and $\kappa = 0.90$ W m$^{-1}$ K$^{-1}$, is a design value (Supplementary Table 14); replacing it by an electrical insulator ($\sigma = 10^{-6}$ S m$^{-1}$, $S = 0$) with $\kappa = 0.3$ or 1.0 W m$^{-1}$ K$^{-1}$ was certified on the same raster. The reference pinwheel of Fig. 2 and Extended Data Fig. 1 uses p and n phases with $\sigma = 10^5$ S m$^{-1}$, $S = \pm 200$ μV K$^{-1}$ and $\kappa = 1.0$ W m$^{-1}$ K$^{-1}$ and a weak phase with $\sigma = 10^3$ S m$^{-1}$, $S = 0$ and $\kappa = 0.2$ W m$^{-1}$ K$^{-1}$. Designs were searched by global and local optimization over continuous geometric parameters with exact p4 symmetry (Supplementary Note 8). Every reported optimum was re-evaluated by fixed-geometry refinement and certified bounds.

**Annulus and winding calculations.** Finite annuli and spiral circuits were solved with nodal finite elements; geometries, the longitudinal correction, the interface model and the measurement settings are given in Supplementary Notes 6 and 8. For a homogeneous annulus of thickness t, azimuthal symmetry gives $j_r = 0$ and a counter-clockwise current $I = -t\sigma_{eff}S_a\Delta T$ with $\Delta T = T_{in} - T_{out} > 0$, independent of the radii, and a central field $B_z(0) = -\mu_0 t\sigma_{eff}S_a\Delta T(r_{in}^{-1} - r_{out}^{-1})/[2\ \ln(r_{out}/r_{in})]$. The certified current and field follow from optimizing $\sigma_{eff}S_a$ over the certified energy-matrix bounds of the Fig. 2b cell. These hold at fixed ΔT. For a prescribed heat input q′ per unit length, the circulating current carries a radial Peltier heat flux, the radial heat flux becomes $-\kappa_{eff}(1 + zT_a)\partial_r T$ with $zT_a = T_0\sigma_{eff}S_a^2/\kappa_{eff}$, and the electromotive force per turn of a closed annulus becomes $S_a q'/[\kappa_{eff}(1 + zT_a)]$. The open spirals carry no circulating current; $zT_a$ is 0.019 for design A and about 0.002 for the pinwheel annulus of Fig. 4.

**Three-dimensional calculations.** Voxel finite volumes in the symmetric form were solved with conjugate gradients[45] preconditioned by smoothed-aggregation algebraic multigrid,[46,47] with periodic boundaries in all directions. Nodal trilinear finite elements on the same voxel geometries served as an independent method. Certified three-dimensional intervals combine the finite-element energy as an upper bound with a complementary bound from divergence-free fluxes reconstructed from the finite-volume face fluxes, a construction that we built on the complementary principle of thermoelectric composites[22] and on scalar periodic bounds of the same kind;[41,42] for W2 with separated junctions the bounds were re-evaluated in ball arithmetic (Supplementary Note 5). All structures with certified intervals have verified four-fold colour symmetry (4, 4/m or a $4_1$ screw axis), which reduces $S_{a,z}$ to a ratio of linear functionals of the $6 \times 6$ matrix, bounded as in two dimensions. Mirror images were obtained by reflecting the whole structure. The three-dimensional cells used the p-type and n-type phases above and, as weak phase, the third phase of design A, except that five straight-rod atlas cells used the carbon composite instead; the insulating-matrix variants are specified separately. Helical stacks of layers related only by a $4_1$ screw axis also carry certified non-zero responses, and their two-phase versions vanish (Supplementary Table 15).

**Colour groups.** For each voxel structure we tested the 16 operations of 4/mmm, combined with every lattice translation of the voxel grid, for invariance of the full material map without exchanging phases, so the colour group is that of the voxel map used in the calculation. In the language of colour symmetry this is the subgroup of operations paired with the identity permutation of colours;[48] for brevity we call its point-group part the colour group. Phase-permuting operations were not considered, because the phases have distinct transport coefficients and no permutation leaves the transport equations invariant. An axial-vector component is allowed when it is unchanged under the average of det(g)g over the invariant operations g, and this test reproduces the 13 pyroaxial point groups.[18]

**Woodpiles and net-derived structures.** Woodpiles, the layer-by-layer rod lattices of photonic crystals[49] that extrusion printing produces as alternating layers of parallel lines,[26] were built from four layers of two rods each, rotated by 90° between layers; segmented rods consist of p-type, n-type and weak segments of unequal length. Nets were taken from the RCSR database[25] and selected in three steps: tetragonal space group (862 of 3,749), at most eight vertices in the conventional cell (231) and $0.5 \leq c/a \leq 1.6$ (126). Nodes and listed edge centres were expanded by the space-group operations, with origin choice 2 where the group has two, and each strut joins the node pair whose midpoint is an edge centre, the pairs chosen so that every node has its coordination number at the smallest total squared strut length. The 116 entries that fix their struts uniquely were computed (Supplementary Note 5), and the struts were voxelized as cylinders. A strut with in-plane orientation φ, measured counter-clockwise from the a axis, was coloured p-type for $0° < \varphi \bmod 90° < 45°$ and n-type for $45° < \varphi \bmod 90° < 90°$; vertical struts and struts at exactly 0° or 45° were p-type, in a weak matrix. Strut radii, voxel counts and junction treatments are given in Supplementary Note 5.

## Data availability

All computed data underlying the figures are archived with the code at Zenodo (https://doi.org/10.5281/zenodo.23668925).

## Code availability

Homogenization, certification and design codes are available at GitHub (https://github.com/joohan-git/antisymmetric-thermopower) and archived at Zenodo (https://doi.org/10.5281/zenodo.23668925, version 1.0.0). The computational scripts used in this study were developed with assistance from Claude Code (Anthropic). All AI-assisted code was reviewed and tested by the author.

## Author contributions

J.N. conceived the study, performed the calculations and wrote the paper.

## Competing interests

The author declares no competing interests.

## Extended Data

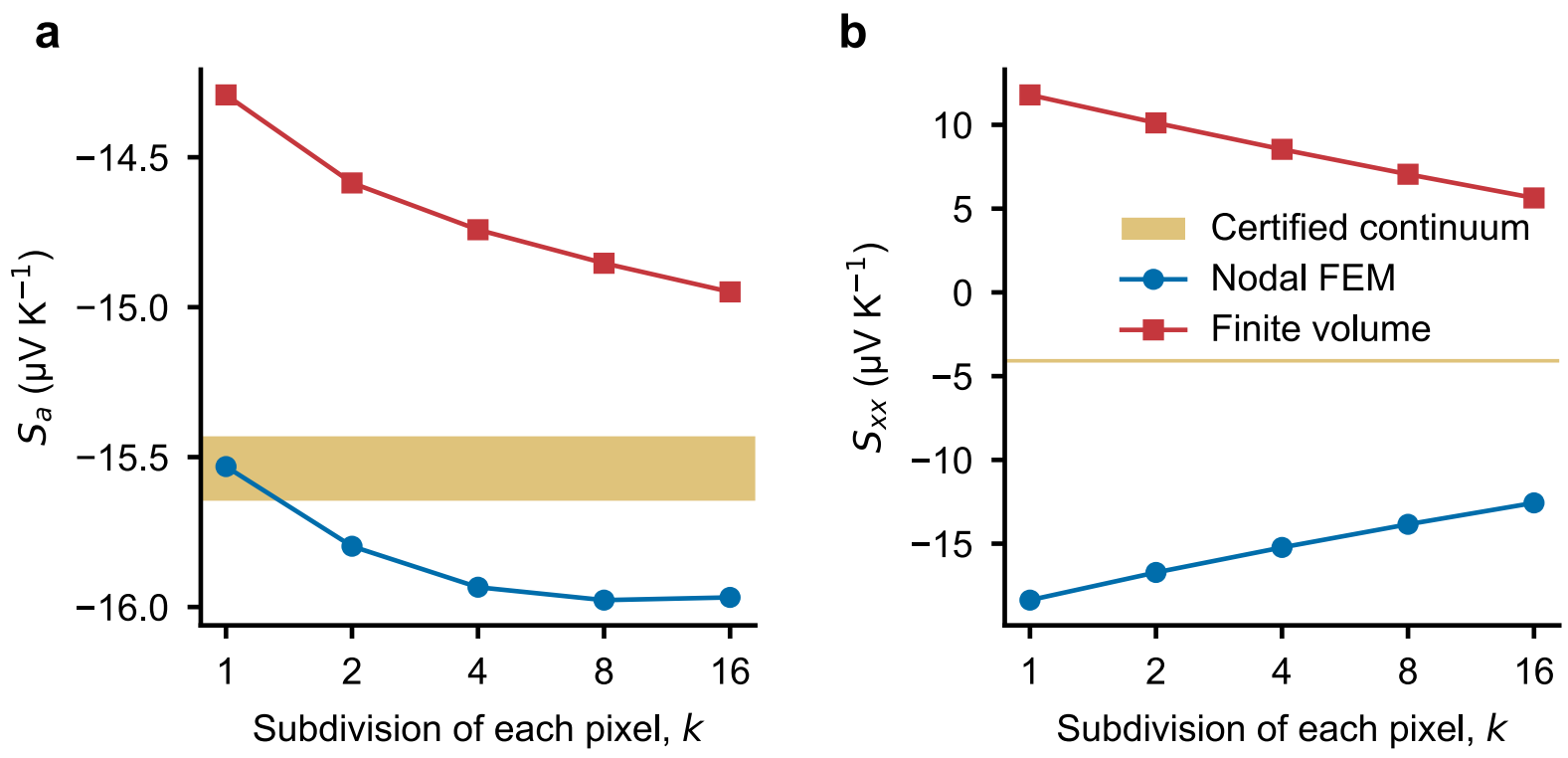

**Extended Data Fig. 1 | Continuum values are certified rather than extrapolated. a**, Axial coefficient $S_a$ and **b**, diagonal coefficient $S_{xx}$ of the 24 × 24 pinwheel raster of Fig. 2b under k-fold subdivision of each pixel, from nodal finite elements (blue) and finite volumes (red). The ochre band is the certified continuum interval from upper and lower bounds on the effective energy matrix. Corner singularities limit the energy convergence of nodal elements asymptotically to about $h^{0.25}$; their $S_a$ moves non-monotonically and remains outside the interval at 16-fold subdivision.

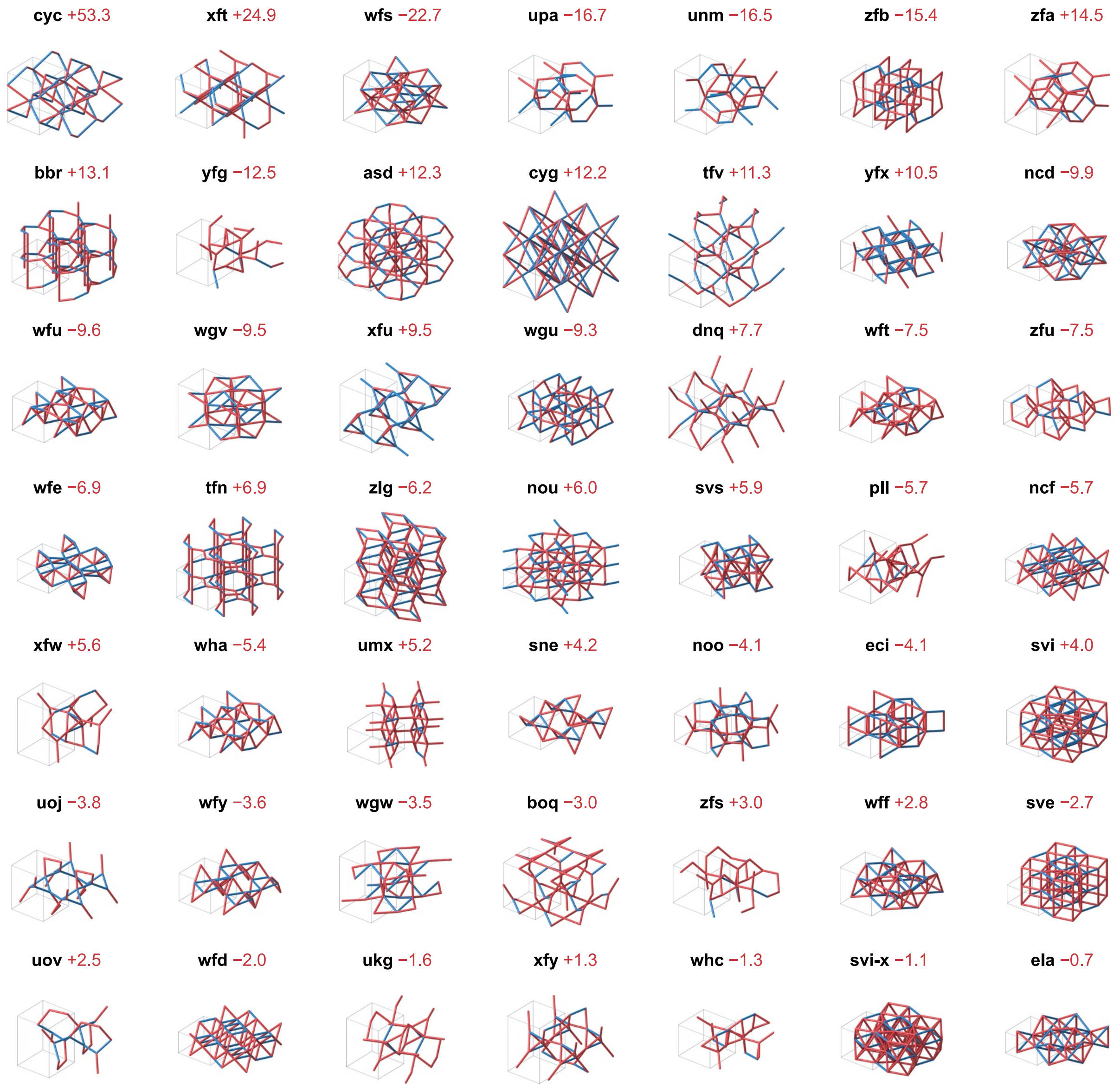

**Extended Data Fig. 2 | Net-derived structures whose colour group allows the axial response.** The 49 structures of Fig. 6 whose colour group allows $S_{a,z}$, ordered by $|S_{a,z}|$, with finite-volume values in μV K$^{-1}$; struts are p-type (red) or n-type (blue) by their in-plane orientation, as in Fig. 6. Struts are drawn with a uniform radius of 0.025a, thinner than the computed 0.08a (Supplementary Note 5), at one common scale; blocks are 2 × 2 × 2 cells, or 2 × 2 × 1 for nets with more than 16 struts per cell or c/a > 1.2, and pieces cut off by the block edge and the weak matrix are not drawn. The thin grey frame marks one unit cell.

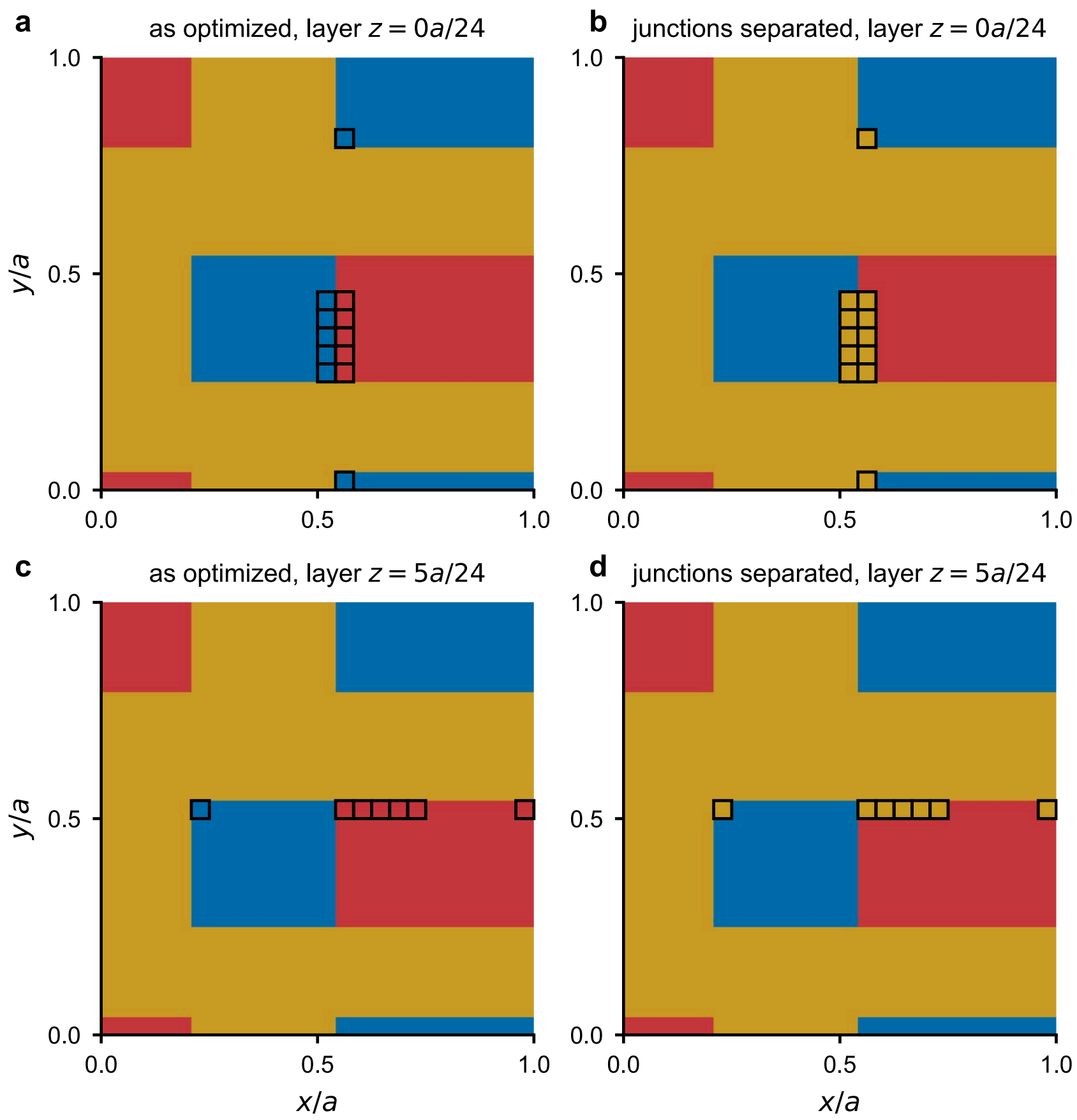

**Extended Data Fig. 3 | Junctions of the woodpile W2 before and after separation.** Two representative layers of the 24 × 24 × 24 W2 cell; the other affected layers follow by successive 90° rotations and translations of a/4 along z. **a**,**c**, As optimized. **b**,**d**, Junctions separated: the outlined voxels, 76 in the cell or 0.55% of its volume, are replaced by the weak phase, opening gaps two voxels (a/12) wide and removing edge-only and vertex-only conductive contacts (Supplementary Note 5). p-type red, n-type blue, weak phase ochre. The certified $S_{a,z}$ = 76–84 μV $K^{-1}$ belongs to the separated geometry; the as-optimized geometry gives 57–87 μV $K^{-1}$ (Fig. 5d).

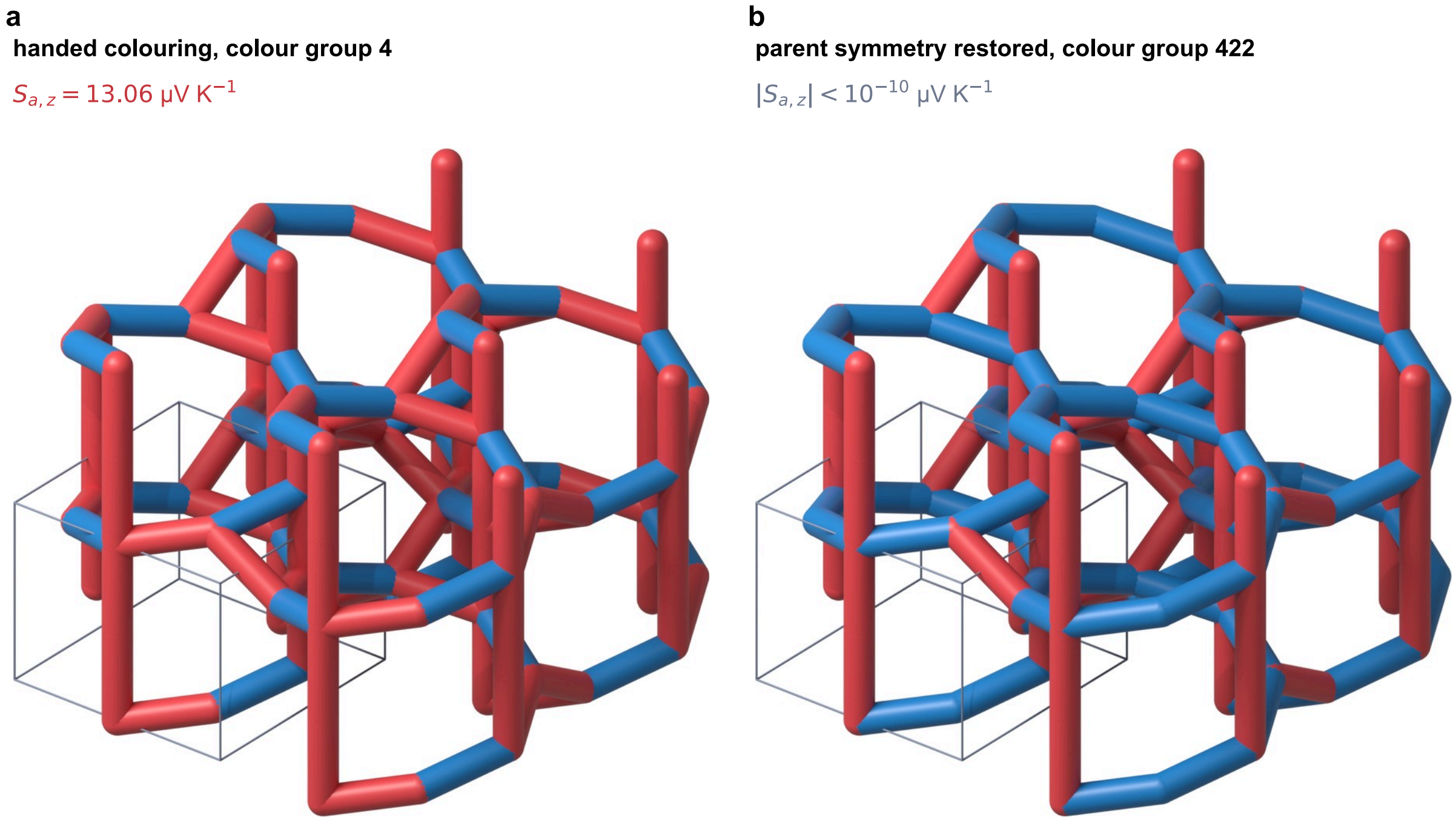

**Extended Data Fig. 4 | Colour symmetry switches the axial response of one net.** The net **bbr** (2 × 2 × 2 cells, struts drawn at radius 0.06a; the weak matrix is not drawn). **a**, Handed colouring by strut orientation modulo 90° (p-type red for 0–45°, n-type blue for 45–90°): colour group 4, $S_{a,z}$ = 13.06 μV K$^{-1}$ (16-voxel finite-volume estimate; certified 2 to 25 μV K$^{-1}$ with the original c/a, Fig. 5d). **b**, The same struts coloured by the folded angle, which restores the parent symmetry: colour group 422, whose two-fold axes perpendicular to z forbid $S_{a,z}$; the computed residual is below $10^{-10}$ μV K$^{-1}$. Geometry and materials are the same; the p/n fractions differ (Supplementary Table 7).

**Supplementary Information**

**Antisymmetric thermopower without magnetism in three-phase composites**

Joohan Nam[1]*

[1]Chair of Inorganic Chemistry I, Technische Universität Dresden, Bergstraße 66, 01069 Dresden, Germany.

*e-mail: joohan.nam@mailbox.tu-dresden.de

**Supplementary Note 1. Effective tensors and the axial coefficient.** The effective scaled energy matrix $C_{eff}$ is 4 × 4 in two dimensions, with rows and columns ordered as (V,x), (V,y), (T,x), (T,y). Its electrical block gives $\sigma_{eff}$, and its off-diagonal block gives $L_{eff}$. Onsager reciprocity requires $C_{eff}$ to be symmetric, so the off-diagonal block satisfies $C^{TV} = (C^{VT})^T$. Reciprocity therefore leaves the spatial symmetry of $L_{eff}$ free, and $S_{eff} = \sigma_{eff}^{-1}L_{eff}$ can carry an antisymmetric part. For a cell with a three-fold or higher rotation axis, $\sigma_{eff}$ is isotropic, and $S_a = (L_{xy} - L_{yx})/(2\sigma_{eff})$.

**Supplementary Note 2. Two-phase reduction.** For two phases with symmetric positive-definite scaled matrices $s_1$ and $s_2$, a congruence transformation W diagonalizes both. The coupled problem then splits into two scalar conduction problems with effective conductivity tensors $\sigma^*(\lambda_1)$ and $\sigma^*(\lambda_2)$.[1–3] Every block of $C_{eff}$ is a linear combination of these two tensors and is therefore spatially symmetric. $S_{eff}$ is the product $\sigma_{eff}^{-1}L_{eff}$ of two symmetric tensors and is symmetric when they commute. A three-fold or higher rotation axis makes both tensors isotropic, so $S_a = 0$. In lower symmetry the principal axes of $\sigma^*(\lambda_1)$ and $\sigma^*(\lambda_2)$ can differ, and a small $S_a$ appears (Supplementary Table 2).

**Supplementary Note 3. Material condition for three phases.** Three phases reduce to two decoupled problems when their scaled matrices lie in a two-dimensional linear space.[3] More generally, exact two-phase results do not carry over to three phases.[4] For any conductivities this occurs exactly when $D = \det[1, S_i, \kappa_i/(T_0\sigma_i) + S_i^2]$ vanishes, since $\det[\sigma_i, \sigma_iS_i, \kappa_i + T_0\sigma_iS_i^2] = T_0\sigma_1\sigma_2\sigma_3D$ (Methods, equation (3)). $S_a$ then vanishes for every geometry with a three-fold or higher rotation axis, and it grows as the third phase is detuned from this condition (Supplementary Fig. 2). Linear independence is necessary but not sufficient for a spatially antisymmetric coupling block. In two dimensions a second exact zero set exists: if $\sigma_1\kappa_1 = \sigma_2\kappa_2 = \sigma_3\kappa_3$, the scaled matrices have equal determinants, and after one common rescaling every phase tensor $s_i \otimes I$ has the form $\Lambda_i \otimes P$ with $\det \Lambda_i \det P = 1$. This set is an exact relation of two-dimensional thermoelectric composites.[5] For the pinwheel of Fig. 2b with its weak phase replaced by $\sigma = 10^3$ S m$^{-1}$, $S = 0$ and $\kappa = 100$ W m$^{-1}$ K$^{-1}$, so that all three phases have $\sigma\kappa = 10^5$ W S m$^{-2}$ K$^{-1}$, the certified interval of $S_a$ contains zero and contracts under refinement (Supplementary Table 17). For a laminate the closure follows because the adjugate of a 2 × 2 matrix is linear, so the harmonic mean of the normal components is proportional to the arithmetic mean of the tangential ones. The full exact-relation theorem gives, after undoing the normalization, an effective matrix $\Lambda^* \otimes P^*$ for every microstructure, so $L_{eff} = \Lambda^*_{12}P^*$ and $\sigma_{eff} = \Lambda^*_{11}P^*$ are proportional and $S_{eff} = (\Lambda^*_{12}/\Lambda^*_{11})I$ is a scalar, whatever the geometry, even when $D \neq 0$. Finite-resolution solvers need not preserve this continuum exact relation. The sweep of Supplementary Fig. 2 stays off this set: its third phase has at least twice the $\sigma\kappa$ of the p-type and n-type phases, which share $\sigma\kappa = 10^5$ S W m$^{-2}$ K$^{-1}$. Equation (11) remains the cubic Taylor coefficient along the straight interpolation path, which leaves the equal-determinant set at intermediate contrast. For weak overall contrast the first-order and second-order corrections to the effective matrix are spatially symmetric, so the first symmetry-allowed axial term is cubic in the contrast. The Methods (equation (11)) give its closed form, derived in Supplementary Note 7: the phase-count determinant multiplying material-weighted three-point correlations of the arrangement; for the pinwheel cell with the midpoint reference the odd part of the axial coupling c scales with exponent 3.0004, and traceless normalized contrasts cancel the cubic term while a quartic term can remain.

**Supplementary Note 4. Corner singularities and certification.** At a vertex where phases meet, the potential behaves as $r^\lambda$ with $\lambda$ set by the phase sequence around the vertex.[6] For the pinwheel cell the slowest exponent is $\lambda = 0.123$, and for the cell with copper it is $\lambda = 0.037$. Design A as drawn contains two point contacts between like phases with $\lambda \approx 0.013$, which widens its certified interval at low enrichment. Its finite-volume point estimates (52.0–52.1 μV K$^{-1}$) lie above the certified interval for the same reason, and the wider interval of Supplementary Table 4 is the same design on a 200 × 200 raster with the lowest enrichment that fitted in memory. Uniform refinement then converges with an energy error of order $h^{2\lambda}$, by analogy with corner-singularity estimates on polygonal domains,[7] which explains the failure of extrapolation in Extended Data Fig. 1. The certified scheme resolves each vertex with singular functions[8] and bounds the effective matrix from both sides (Supplementary Fig. 3). Refining the angular and radial enrichment (m = 2, 4, 8) tightens the enclosure monotonically. The coupled angular problem, $\int Y'^{T}sZ'\,d\theta = \lambda^2\int Y^{T}sZ\,d\theta$ with continuity of Y and of $sY'$ across interfaces, is self-adjoint in the s-weighted inner product, so $\lambda^2$ is real and non-negative and repeated eigenvalues need no logarithmic modes. All modes of the discretized angular problem (m = 2–8 elements per phase sector) with $\lambda > 0$ and the two constant modes were kept; negative powers and logarithms carry infinite energy and were excluded. Discs of radius 0.22 pixel were placed at the vertices, and the surrounding mapped bilinear elements share the full trace basis with the disc functions on the circle, so the trial space is $H^1$-conforming (trace mismatch $2 \times 10^{-15}$). The interval limits, roots of the support-function equations of the Methods, were found by bracketing and Brent iteration without reducing the matrix constraint to component bounds.

**Supplementary Note 5. Three-dimensional certification and colour groups.** In three dimensions the two-dimensional duality is not available. The conforming nodal finite-element energy bounds the effective energy matrix from above. Face fluxes of the finite-volume solution, reconstructed as lowest-order Raviart–Thomas fields[9] in each voxel and projected to zero divergence, give a complementary upper bound on the effective resistance matrix.[3,10,11] Inverting the latter bounds the energy matrix from below, and the axial coefficient is then bounded over the enclosed set under the verified four-fold symmetry (Supplementary Table 8). Where like phases meet only along voxel edges, the two point estimates differ most; regularizing these junctions narrows both the interval and the gap (Supplementary Table 9). Colour groups were determined by testing all operations of 4/mmm with lattice translations, so the listed groups are the colour-preserving subgroups of 4/mmm; operations outside 4/mmm were not searched, and the allowed verdicts rest also on the computed non-zero $S_{a,z}$, and the counts by family are listed in Supplementary Table 6; every computed net is listed in Supplementary Table 16. **Computational details.** In each voxel the face fluxes were interpolated linearly along each axis (the tensor-product extension of the lowest-order Raviart–Thomas element on rectangles[9]), and a periodic discrete Hodge projection removed the residual divergence of the iterative solution. The complementary energy of each component was integrated exactly, and the resulting resistance matrix was inverted to give the lower bound on the energy matrix. For W2 with separated junctions the bounds were also evaluated with ball arithmetic (Arb): dyadic conforming potentials and exactly divergence-free fluxes were integrated with integer polynomial moments, and semidefinite dual witnesses bounded $S_{a,z}$. The verification used 192-bit Arb arithmetic (python-flint), nodal values fixed as integers divided by $2^{40}$, exact rational material values and positive-definiteness checks of the witnesses by Arb $LDL^T$ factorization. On the twofold-refined voxels it gives 74.37–85.92 μV $K^{-1}$ (rounded outward), the same interval as the floating-point bounds to within $3 \times 10^{-8}$ μV $K^{-1}$; the narrower 76–84 μV $K^{-1}$ is the floating-point enclosure on threefold-refined voxels. They apply to the stated voxel geometries. In-plane conductivities of the woodpiles are about $5 \times 10^3$ S $m^{-1}$. For the insulating matrix ($\sigma = 10^{-6}$ S $m^{-1}$, $S = 0$, $\kappa = 0.3$ or 1.0 W $m^{-1}$ $K^{-1}$), face connectivity of the p and n voxels was tested on the periodic label graph. The straight-rod woodpiles of the atlas used two of the p-type, n-type and weak phases as rods, with the third phase or the carbon composite as matrix. Nets were built from the NODE and EDGE_CENTER entries of the RCSR file. Nodes and edge centres were expanded with origin choice 2 for the tetragonal groups that have two origin choices; in this setting every edge centre of the computed entries is the midpoint of a node pair. Each strut joins the node pair (with its lattice translation) whose midpoint is an edge centre, one pair per centre, chosen so that every node has its coordination number and the total squared strut length is smallest; the resulting strut set was checked to be invariant under the space group. Of the 126 entries that pass the other filters, ten do not fix their struts in this way and were not computed: in srs-c4*, cao, tej, dia-b3**, trn-c and pts-y some edge centres are the midpoint of no node pair, in keb and dia-c4** the listed centres are fewer than the coordination numbers require, and in dia-c4 and sql-c* several assignments have the same smallest length; a symmetry filter would select one, but these entries were excluded so that every computed structure follows from its entry without a tie-break. Strut radii were 0.08a for every net of the atlas and, for the certified structures, 0.08a (**cyc**, **bbr**) and 0.12a (**unm**), with 16 voxels along a and b and $N_z = 4\lceil \text{round}(16c/a)/4 \rceil$ along c; woodpile W1 used 16 and W2 24 voxels per edge. The 82 segmented woodpiles of the atlas have random segment parameters on 16-voxel cells; W1 is the one with the second-largest $|S_{a,z}|$ (45 μV $K^{-1}$), and W2 was obtained from W1 by Nelder–Mead optimization of the same parameters on 24-voxel cells, so it lies outside the atlas. Certified net values use rectangular voxels that restore the original c/a ratios (1.414 for **cyc**, 1.468 for **unm** and 0.770 for **bbr**). Systre reports collisions in the barycentric embedding of **cyc**; the RCSR coordinates used here place every node at a distinct position. In the structure derived from **cyc** the n-type struts form isolated islands and the p-type struts percolate along z, while the p–n union percolates in all three directions. Separated woodpile junctions replace the like-phase voxels on both sides of every edge or vertex contact by the weak phase; in W2 the four crossings where weak, n-type, weak and p-type segments meet were also cleared to the weak phase. Fused junctions instead fill the adjacent weak voxels with the contacting phase. All modified structures keep their colour group, and every structure was refined at fixed geometry by subdividing each voxel k-fold.

**Supplementary Note 6. Fabrication tolerance, interface resistance and finite samples.** Junction separation, corner rounding and random boundary errors were applied to design A on fixed rasters (Supplementary Table 4). Random errors that break the p4 symmetry can lower $S_a$ substantially, but protecting a 4% separation at the two junctions removes these failures. Interface resistance, with contact resistivities spanning and exceeding those of metal–$Bi_2Te_3$ junctions,[12] lowers $\sigma_{eff}$ much more than $S_a$ (Supplementary Table 11). For finite samples, a short probe at the same height and phase on both sides gives the effective-medium voltage (Supplementary Table 13). Two-phase p4 rings with a conductivity contrast of 100 carry boundary-sensitive spurious currents with eight cells, 1.1–1.2 mA $K^{-1}$ for the unshifted tiling and up to 1.4 mA $K^{-1}$ when the cell origin is shifted by a quarter cell, several times the 0.26–0.28 mA $K^{-1}$ of the pinwheel annulus of Fig. 4. Lowering the contrast to 3 and adding homogeneous borders reduces them, with 16 or more cells, to 0.1–2% of the current of a homogeneous design-A annulus (1.6 mA $K^{-1}$). Mirror-symmetric rings, which give below $10^{-13}$ A, are therefore the preferred null control.

**Supplementary Note 7. Derivation of the leading axial thermopower.** This note derives equation (11) of the Methods from the cell problem. Each phase i has the symmetric 2 × 2 transport matrix $s_i$ of equation (1), and the local conductivity is $K(\mathbf{x}) = s(\mathbf{x}) \otimes I$, acting on the gradients ($\nabla V$, $\nabla T$) ordered ($V_x$, $V_y$, $T_x$, $T_y$). For a macroscopic gradient $\mathbf{E}$, the periodic corrector $\varphi$ solves

$$\nabla\cdot[K(\mathbf{x})(\mathbf{E} + \nabla\varphi)] = 0, \quad C\mathbf{E} = \langle K\mathbf{e}\rangle, \quad \mathbf{e} = \mathbf{E} + \nabla\varphi, \tag{S1}$$

where $\langle\cdot\rangle$ is the cell average. Splitting $K = K_0 + \delta K$ with a homogeneous reference $K_0 = s_0 \otimes I$ and writing the equation in Fourier space, with $\mathbf{k}$ standing for $I_2 \otimes \mathbf{k}$, the gradient acting on both potentials, $\mathbf{k}^T(K_0\hat{\mathbf{e}}) = -\mathbf{k}^T(\delta K\mathbf{e})^\wedge$ with $\hat{\mathbf{e}}(\mathbf{k})$ parallel to $\mathbf{k}$ for $\mathbf{k} \neq 0$, gives the Lippmann–Schwinger equation

$$\mathbf{e} = \mathbf{E} - \Gamma(\delta K\mathbf{e}), \quad \hat{\Gamma}(\mathbf{k}) = \mathbf{k}(\mathbf{k}^T K_0 \mathbf{k})^{-1}\mathbf{k}^T = s_0^{-1} \otimes \hat{\mathbf{k}}\hat{\mathbf{k}}^T \ (\mathbf{k} \neq 0), \tag{S2}$$

with $\hat{\Gamma}(0) = 0$, because $\mathbf{k}^T K_0 \mathbf{k} = |\mathbf{k}|^2 s_0$ for the isotropic reference. Iterating and using $\langle K_0\Gamma\boldsymbol{\tau}\rangle = 0$ for every periodic field $\boldsymbol{\tau}$, because $\hat{\Gamma}(0) = 0$, gives the weak-contrast expansion of periodic homogenization,[3,13]

$$C = K_0 + \langle \delta K \rangle - \langle \delta K \Gamma \delta K \rangle + \langle \delta K \Gamma \delta K \Gamma \delta K \rangle - \ldots \tag{S3}$$

With cell-normalized Fourier coefficients, $\langle fg \rangle = \Sigma_k\, \hat{f}(-\mathbf{k})\hat{g}(\mathbf{k})$. The axial coefficient is $c = (C_{Vx,Ty} - C_{Vy,Tx})/2$, the spatially antisymmetric part of the V–T block. The first-order term $\langle \delta s \rangle \otimes I$ is isotropic. The second-order term is

$$\langle \delta K \Gamma \delta K \rangle = \Sigma_{k \neq 0}\, \delta\hat{s}(-\mathbf{k}) s_0^{-1} \delta\hat{s}(\mathbf{k}) \otimes \hat{\mathbf{k}}\hat{\mathbf{k}}^T; \tag{S4}$$

every spatial factor $\hat{\mathbf{k}}\hat{\mathbf{k}}^T$ is symmetric, so this term adds nothing to c for any geometry and any materials. The third-order term is

$$\langle \delta K \Gamma \delta K \Gamma \delta K \rangle = \Sigma_{k,\, q \neq 0}\, M(\mathbf{k}, \mathbf{q}) \otimes (\hat{\mathbf{k}}\cdot\hat{\mathbf{q}})\hat{\mathbf{k}}\hat{\mathbf{q}}^T, \quad M(\mathbf{k}, \mathbf{q}) = \delta\hat{s}(-\mathbf{k}) R \delta\hat{s}(\mathbf{k} - \mathbf{q}) R \delta\hat{s}(\mathbf{q}), \quad R = s_0^{-1}, \tag{S5}$$

whose spatial factor is not symmetric. Its antisymmetric part is $\tfrac{1}{2}(\hat{\mathbf{k}}\cdot\hat{\mathbf{q}})(\hat{k}_x\hat{q}_y - \hat{k}_y\hat{q}_x)\, J$ with $J = [0, 1; -1, 0]$, so

$$c^{(3)} = \tfrac{1}{2}\, \Sigma_{k,\, q \neq 0}\, [M(\mathbf{k}, \mathbf{q})]_{VT}\, w(\mathbf{k}, \mathbf{q}), \quad w(\mathbf{k}, \mathbf{q}) = (\hat{\mathbf{k}}\cdot\hat{\mathbf{q}})(\hat{\mathbf{k}}\times\hat{\mathbf{q}})\cdot\hat{z}. \tag{S6}$$

The relabelling $(\mathbf{k}, \mathbf{q}) \to (-\mathbf{q}, -\mathbf{k})$ leaves the sum unchanged, gives $w(-\mathbf{q}, -\mathbf{k}) = -w(\mathbf{k}, \mathbf{q})$ and, because every $\delta\hat{s}(\mathbf{k})$ is a symmetric $2 \times 2$ matrix, $M(-\mathbf{q}, -\mathbf{k}) = M(\mathbf{k}, \mathbf{q})^T$. Averaging the two forms shows that only the antisymmetric part of M contributes: $c^{(3)} = \tfrac{1}{4}\, \Sigma\, [M - M^T]_{VT}\, w$. For phases with indicator functions $\chi_i$, $\delta s = \Sigma_i\, \chi_i \Delta_i$ gives

$$c^{(3)} = \tfrac{1}{2}\, \Sigma_{ij\ell}\, [\Delta_i R \Delta_j R \Delta_\ell]_{VT}\, G_{ij\ell}, \quad G_{ij\ell} = \mathrm{Re}\, \Sigma_{k,\, q \neq 0}\, \hat{\chi}_i(-\mathbf{k})\hat{\chi}_j(\mathbf{k} - \mathbf{q})\hat{\chi}_\ell(\mathbf{q})\, w(\mathbf{k}, \mathbf{q}), \tag{S7}$$

with $G_{ij\ell} = -G_{\ell ji}$ from the same relabelling; in particular $G_{iji} = 0$. Only the part of the material product that is antisymmetric under reversal of its factors survives, because the reversed product is the transpose. Take phase 3 as the reference, $s_0 = s_3$, so that only $\chi_1$ and $\chi_2$ appear and $\Delta_i = s_i - s_3$. With the interpolation $s = s_3 + \varepsilon(\chi_1\Delta_1 + \chi_2\Delta_2)$ of the Methods, the n-th term of the series is of order $\varepsilon^n$, and $c^{(3)} = \varepsilon^3 c_3$ with $c_3$ of Methods equation (9). With $Q = s_3^{-1/2}$ write

$$Q\Delta_i Q = \theta_i I + d_{i,z}\tau_z + d_{i,x}\tau_x, \quad \tau_z = [1, 0; 0, -1], \quad \tau_x = [0, 1; 1, 0], \quad \theta_i = \tfrac{1}{2}\,\mathrm{tr}(s_3^{-1}s_i) - 1. \tag{S8}$$

Then $\Delta_i R \Delta_j R \Delta_\ell = Q^{-1}(D_i D_j D_\ell)Q^{-1}$ with $D_i = Q\Delta_i Q$, and for any $2 \times 2$ matrix Y the antisymmetric part of $Q^{-1}YQ^{-1}$ is $(\det s_3)^{1/2}$ times that of Y. Because $\tau_z\tau_x = J = -\tau_x\tau_z$ and $\tau_z^2 =$

$\tau_x^2 = I$, the J component of a triple product is

$$D_iD_jD_\ell|_J = \theta_i(\mathbf{d}_j\times\mathbf{d}_\ell) + \theta_j(\mathbf{d}_i\times\mathbf{d}_\ell) + \theta_\ell(\mathbf{d}_i\times\mathbf{d}_j), \quad \mathbf{d}\times\mathbf{d}' = d_zd'_x - d_xd'_z. \tag{S9}$$

With two indices taking the values 1 and 2 and $G_{iji} = 0$, the non-zero terms are $G_{112} = -G_{211}$ and $G_{122} = -G_{221}$, with J components $2\theta_1(\mathbf{d}_1\times\mathbf{d}_2)$ and $2\theta_2(\mathbf{d}_1\times\mathbf{d}_2)$; the partner terms 211 and 221 contribute equally. Because $J_{12} = 1$, the V–T element of the antisymmetric part equals its J component, and with the factor ½ of (S7) and the factor $(\det s_3)^{1/2}$ of the congruence

$$c_3 = 2(\det s_3)^{1/2}(\mathbf{d}_1\times\mathbf{d}_2)(\theta_1G_{112} + \theta_2G_{122}). \tag{S10}$$

The prefactor is proportional to the phase-count determinant. The congruence $X \to Q^{-1}XQ^{-1}$ is a linear map of the three-dimensional space of symmetric matrices with determinant $(\det s_3)^{3/2}$ in the coordinates $(X_{11}, X_{12}, X_{22})$, and it maps $I + D_i$ to $s_i$ and $I$ to $s_3$. Evaluating the 3 × 3 determinant of the entries of $I + D_1$, $I + D_2$ and $I$ gives $2(\mathbf{d}_1\times\mathbf{d}_2)$, hence

$$2(\det s_3)^{1/2}(\mathbf{d}_1\times\mathbf{d}_2) = \det[s_{i,11}, s_{i,12}, s_{i,22}]_{i=1,2,3}/\det s_3 = \Delta_{123}/(T_0 \det s_3), \tag{S11}$$

where the last step uses $s_{i,22} = (\kappa_i + T_0\sigma_iS_i^2)/T_0$. With a three-fold or higher rotation axis, $\boldsymbol{\sigma}_{\mathrm{eff}} = aI$ with $a = s_{3,11} + O(\varepsilon)$, and $S_a = c/a$ gives equation (11) of the Methods. The same steps hold in three dimensions with $\hat{\mathbf{k}}\times\hat{\mathbf{q}}$ kept as a vector; its z component gives $c_z^{(3)}$, and $S_{a,z} = c_z/a$ for cells with a three-fold or higher axis along z. Under an orthogonal transformation O the vector transforms as $(\det O)O\mathbf{G}$, so a mirror reverses its components parallel to the mirror plane and preserves its normal component. The result has the following properties. (i) The axial part of C is zero at first and second order for every geometry and every set of materials. (ii) For two phases only one indicator remains, so $G_{111} = 0$ and $c^{(3)} = 0$; two phases decouple at all orders,[1,2] so every block of C stays symmetric. This makes $S_a$ vanish only when $\sigma_{\mathrm{eff}}$ and $L_{\mathrm{eff}}$ commute, as under a three-fold or higher axis; in lower symmetry a two-phase $S_a$ can be non-zero (Supplementary Note 2). (iii) Linearly dependent transport matrices give $\Delta_{123} = 0$. (iv) In two dimensions a reflection reverses $\hat{\mathbf{k}}\times\hat{\mathbf{q}}$ and hence every G; in three dimensions it reverses the components of **G** parallel to the mirror plane. Arrangements invariant under a reflection that preserves every phase label therefore give $G = 0$ in two dimensions, and $G_z = 0$ for mirrors containing z. (v) If $\theta_1 G_{112} + \theta_2 G_{122} = 0$ the cubic term vanishes and $S_a = O(\varepsilon^4)$; in the tested cases with non-commuting traceless contrasts a quartic term remained, whereas commuting traceless contrasts have $\Delta_{123} = 0$ and fall under (iii). Geometric coefficients multiplying ordered products of contrast matrices appear at every order of the corresponding expansion for coupled linear response, which also shows that coefficients antisymmetric in the order of the factors vanish for two phases;[13] three-point correlation parameters of the microstructure likewise enter bounds on the transport coefficients of two-phase composites.[14] An antisymmetric effective conductivity also arises in the classical Hall effect of composites, but there it is inherited from antisymmetric constituent tensors, and its reduction to a symmetric problem is available only for two components.[15] **Numerical checks.** For the reference pinwheel of Fig. 2b (phase 1 p-type, phase 2 n-type, phase 3 the weak reference) we evaluated G by FFT collocation on grids refined 128-fold, a 3072 × 3072 grid, which gave $G_{112} = -6.7630 \times 10^{-4}$ and $G_{122} = 5.91112 \times 10^{-3}$. Extrapolations of the collocation sums and of truncated sums with exact pixel Fourier coefficients and symmetric cutoffs agree in $G_{112} + G_{122}$ to within a relative $3 \times 10^{-7}$. For the phase-3 reference, linear extrapolation from $\varepsilon = 3 \times 10^{-4}$ and $10^{-4}$ at four-fold refinement gave $S_a/\varepsilon^3 = -1.01308 \times 10^6$ μV K$^{-1}$ with finite volumes. The spectral prediction on that grid was $-1.01663 \times 10^6$ μV K$^{-1}$; the 0.35% difference includes spatial discretization error. For the midpoint reference $s_0 = (s_1 + s_2)/2$ with $s_i(\varepsilon) = s_0 + \varepsilon(s_i - s_0)$, a different interpolation path in which all three phases carry contrast, the cubic coefficient from equation (9) with $R = s_0^{-1}$, its geometric correlations extrapolated to the continuum, is $-0.160870$ μV K$^{-1}$; the odd part of c in the full solution scales with exponent 3.0004 on a 192 × 192 grid over $\varepsilon = 0.025$–$0.00625$, and the extrapolated finite-volume and collocation cubic coefficients differ by a relative $1.4 \times 10^{-6}$. These are numerical extrapolations, not certified bounds. For the segmented woodpile W2 on a $24^3$ grid with dimensionless test matrices $s_3 = [1, 0.3; 0.3, 0.8]$, $\Delta_1 = [0.7, -0.2; -0.2, 0.1]$ and $\Delta_2 = [-0.3, 0.5; 0.5, 0.9]$, full solutions gave $c_z/\varepsilon^3 = 2.34793 \times 10^{-4}$ and $2.36060 \times 10^{-4}$ at $\varepsilon = 0.01$ and $0.005$, approaching the prediction $2.37335 \times 10^{-4}$. Reflection in x reversed the sign. The two-dimensional two-phase and p4mm cells gave zero cubic response, and tested noncommuting traceless contrasts gave a nonzero quartic term. Convergence depends on the reference matrix, not on the conductivity ratio alone. A sufficient condition is $|\varepsilon|\max_i\|s_0^{-1/2}(s_i - s_0)s_0^{-1/2}\|_2 < 1$. The maximum is 0.99 for the midpoint reference, so its Neumann series converges at $\varepsilon = 1$. Along this path the finite-volume $S_a$ follows the cubic term within 5% up to $\varepsilon \approx 0.04$ and exceeds it about 90-fold at $\varepsilon = 1$, where higher orders dominate (Supplementary Fig. 4); convergence of the series does not make its cubic truncation accurate. Here we specialize the third-order skew invariant of coupled-field composites[13] to thermoelectric transport with symmetric isotropic constituents and reduce its material factor to the phase-count determinant weighted by the trace contrasts $\theta_i$. In a response matrix without spatial indices such a term is excluded by symmetry of the matrix;[13] here it multiplies the spatially antisymmetric J, so C remains symmetric.

**Supplementary Note 8. Materials, design search, annulus, winding and measurement calculations. Materials.** The p-type and n-type values of the Methods are model values representative of room-temperature measurements on bulk alloys,[16,17] with κ of the n-type phase taken at the upper end of the reported range. Printed alloys reach about half this conductivity at lower κ,[18] which scales $\sigma_{eff}$ but not the symmetry argument. The Seebeck coefficient of the carbon composite is typical of metallic carbon-nanotube networks.[19] **Design search.** Designs were searched by differential evolution,[20] Sobol sampling[21] and multistart Nelder–Mead[22] over continuous geometric parameters, with exact p4 symmetry enforced; differential evolution and Nelder–Mead used SciPy.[23] **Annulus and winding circuits.** Finite annuli were meshed with nodal finite elements and isothermal inner and outer boundaries. Annuli lay in a 20 × 20 mm box and were represented by the Cartesian bilinear cells whose centres fall between the radii, giving staircase boundaries. The homogeneous annulus on the same mesh carries the effective tensor of the 24 × 24 bilinear cell problem without corner enrichment, the discretization with which every cell of the microstructured annulus is resolved: $\sigma_{eff}$ = 19,736 S m$^{-1}$ and $S_a$ = −15.53 μV K$^{-1}$. Unresolved corner singularities make this discretization overestimate $\sigma_{eff}$; the certified tensor of the same cell has $\sigma_{eff}$ = 16,623–16,634 S m$^{-1}$ and $S_a$ = −15.65 to −15.43 μV K$^{-1}$, and the certified current of 2.56–2.61 × 10$^{-4}$ A K$^{-1}$ follows from it. The two predictions therefore differ by the effective tensor, not by the annulus calculation: on a 576 × 576 grid the homogeneous annulus reproduces the closed forms for its tensor to 10$^{-9}$ for the current and 0.06% for the field. The circulating current and the magnetic moment were integrated from the computed current density. Winding circuits were spiral strips 0.4 mm wide, described by the homogenized tensor of design A, running from 3 to 8 mm radius in an electrically insulating matrix, around an isothermal line source of 2 mm radius inside an isothermal outer boundary at 9.3 mm. Because the homogenized problem has no intrinsic length, the calculation is scale-free, and the strip width stands for any width that spans many cells. The corrected voltage is $V_{corr} = V_{start} - V_{end} - S_{xx}(T_{end} - T_{start})$, where start and end are the inner (3 mm, hotter) and outer (8 mm) terminals and $n > 0$ denotes counter-clockwise winding, with $S_{xx}$ = 8.47 μV K$^{-1}$ (certified 8.16–8.79 μV K$^{-1}$), $S_a$ = 49.8 μV K$^{-1}$ and $\sigma_{eff}$ = 3.22 × 10$^{4}$ S m$^{-1}$. **Interface resistance.** Interface resistance was modelled as a zero-thickness layer with contact resistivity $\rho_c$ = 0–10$^{-6}$ Ω m$^2$ and Kapitza resistance[24] $R_K$ = 0 or 10$^{-5}$ m$^2$ K W$^{-1}$, for cells of 1 mm and 0.3 mm. The upper value of $\rho_c$ exceeds the contact resistivities reported for metal–$Bi_2Te_3$ junctions[12] and inferred for printed legs.[18] The same model gives the two-phase p4 pinwheel of Fig. 2a an axial response that reverses in its mirror image (Supplementary Table 1): a resistive p/n interface is a third transport matrix, so two-phase cells are a null test only with ideal contacts. Kapitza resistance alone gives $|S_a|$ below 1.3 μV K$^{-1}$. **Orthogonal-gradient measurement.** Orthogonal-gradient measurements used square specimens of N × N cells driven once along x and once along y, with isothermal hot and cold faces, adiabatic and electrically insulating other faces and current-free point probes at mid-height; the results, including tilted and shifted probes and shifted terminations, are listed in Supplementary Table 12. **Thermal mismatch.** The winding law assumes that the strip does not perturb the temperature field. With the same strip, source and endpoints, a matrix-to-strip thermal-conductivity ratio of 0.003 gives corrected voltages of 0.94, 0.89 and 0.91 times $nS_a q'/\kappa_{eff}$ for $n = 1$, 2 and 3, and a ratio of 10 gives 0.95 for $n = 2$ (1024 × 1024 grid, point estimates). In all four cases the corrected voltage agrees within 1.1% with $S_a\int(\partial_y T\,dx - \partial_x T\,dy)$ along the strip; thermal-field distortion therefore accounts for most of the winding-law deviation. **Sensor layout and idealized materials.** For design A, $S_{xx} = 8.1$–$8.8\ \mu V\ K^{-1}$; returning along a mirror-image spiral wound in the opposite sense cancels the longitudinal term and doubles the winding signal, with both terminals on one isotherm. With idealized materials outside the Wiedemann–Franz relation, $|S_a|$ exceeds the largest constituent $|S_i|$ by 20% (Supplementary Table 18).

**Supplementary Note 9. Why reciprocal elastic diffusion excludes axial thermopower in uniaxial crystals.** For reciprocal, energy-diagonal (elastic) electronic scattering, the electrical and thermoelectric conductivities are moments of one symmetric transport distribution $\Sigma(E)$: $\sigma = \int\Sigma(E)w(E)dE$ and $L = (q_c T)^{-1}\int(E - \mu)\Sigma(E)w(E)dE$, where $w = -\partial f/\partial E$ and $q_c$ is the signed electronic charge. This includes several bands and anisotropic relaxation times that respect the crystal symmetry; the Mott formula is its low-temperature limit. A three-fold or higher axis makes every symmetric tensor scalar in the plane, so $\sigma$ and $L$ are both scalar there and $S = \sigma^{-1}L$ has no axial part. A non-symmetric tensor invariant under the same rotation could contain $aI + bJ$; the term $bJ$ is excluded by this transport model, not by the rotation. In monoclinic and triclinic crystals $\sigma$ and $L$ remain symmetric but need not commute, and $S - S^T = \sigma^{-1}[L, \sigma]\sigma^{-1}$ can be non-zero.[25] In a point group that permits an axial thermopower, energy-mixing collisions or phonon drag can invalidate this transport-distribution argument; their contribution is not evaluated here. The composite homogenizes locally equilibrated voltage and temperature fields, conserving total charge and heat rather than each electronic energy channel. Reciprocity permits a non-symmetric $L_{eff}$, which excludes any scalar-weighted sum of symmetric spatial tensors, including a common symmetric transport distribution. Two isotropic phases decouple exactly and keep $L_{eff}$ symmetric;[2,3] three linearly independent phase matrices remove this obstruction (Supplementary Note 3).

**Supplementary Note 10. The phase-count rule beyond thermoelectricity.** The argument concerns the coupled-potential problem $j_\alpha = -\Sigma_\beta\, s_{\alpha\beta}(\mathbf{x})\nabla u_\beta$, $\nabla\cdot j_\alpha = 0$ ($\alpha, \beta = 1, 2$), with piecewise constant, symmetric positive-definite $s_i$, spatially isotropic local blocks and continuity of both potentials and both normal fluxes. Simultaneous congruence diagonalization then makes every effective spatial block symmetric for two phases,[2] and a three-fold or higher axis eliminates the transverse antisymmetric part of the zero-flux conversion tensor $(C^*_{11})^{-1}C^*_{12}$, the analogue of the Seebeck tensor. For three phases, the determinant $\det[s_{i,11}, s_{i,12}, s_{i,22}]$, the cubic expansion of Supplementary Note 7 and the phase-preserving axial-vector selection rule apply unchanged when both potentials are ordinary scalars; linear independence is necessary, not sufficient, and the two-dimensional equal-determinant exact relation remains an exception.[3,5,13] The generality of the two-phase reduction is established;[2] what transfers from this work is the axial construction, its determinant condition and the colour-symmetry test. Candidate systems include thermodiffusion and heat–moisture diffusion written with conserved energy and species fluxes and their conjugate forces, and electrokinetics after reduction to a closed, isotropic Darcy-scale law; additional independent fields, interfacial transfer laws or pore-scale hydrodynamics require separate analysis. Static two-potential problems also fall within the mathematical framework,[2] but magnetoelectricity requires the polar–axial symmetry of its fields, and ordinary piezoelectricity does not share the two-scalar-field structure. The transport-distribution argument of Supplementary Note 9 is specific to electronic thermoelectricity and is not part of this generalization.

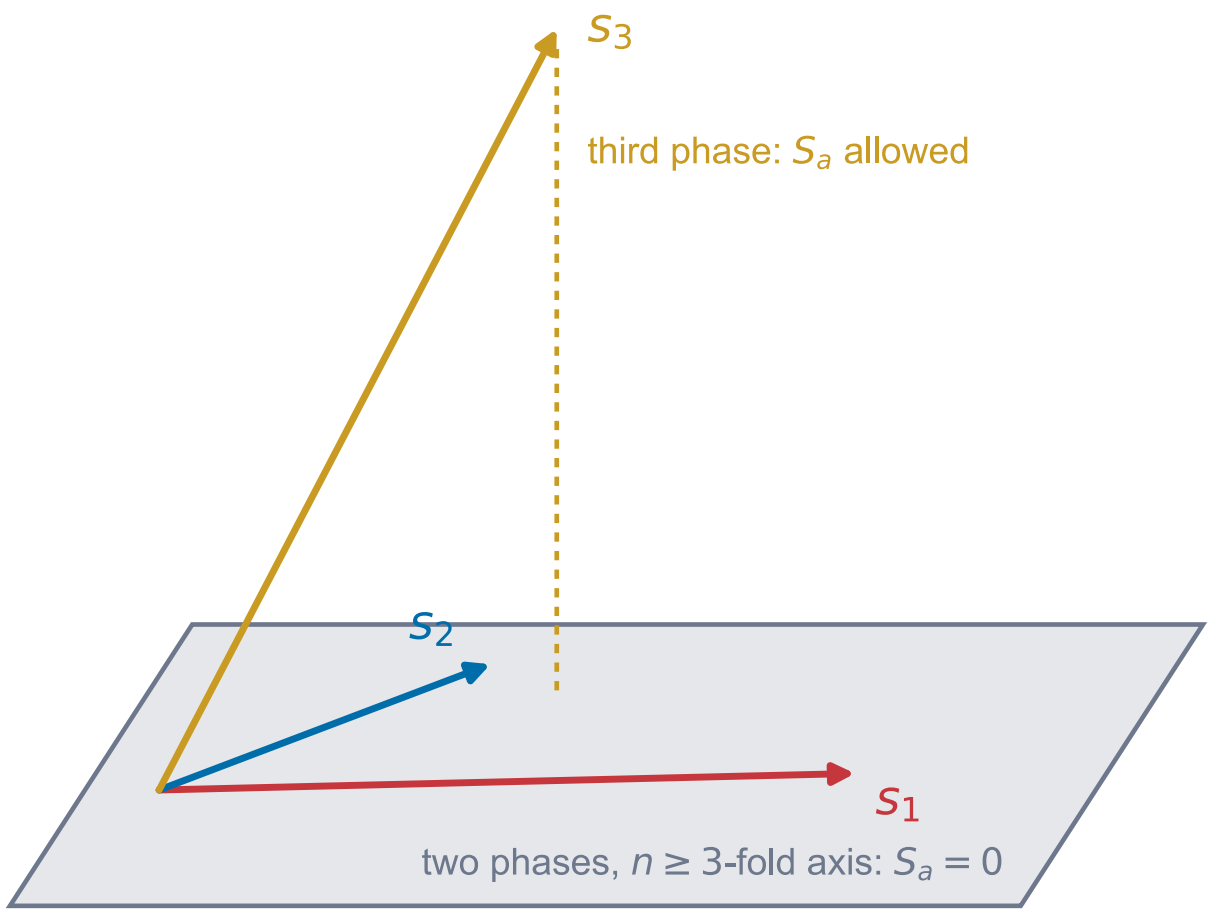

**Supplementary Fig. 1 | Phase-count rule.** The scaled transport matrices $s_i$ of the phases are vectors in the three-dimensional space of symmetric 2 × 2 matrices. Two phases span a plane, and one congruence transformation decouples them, so $\mathbf{L}_{eff}$ is symmetric and, under a three-fold or higher axis, $S_a = 0$ for every geometry. A third phase outside this plane, $\det[\sigma_i, \sigma_i S_i, \kappa_i + T_0 \sigma_i S_i^2] \neq 0$, couples charge and heat irreducibly, and in a ferroaxial-type arrangement $S_a$ is then generically non-zero, except, in two dimensions, when all phases share one product σκ (schematic; Supplementary Note 3).

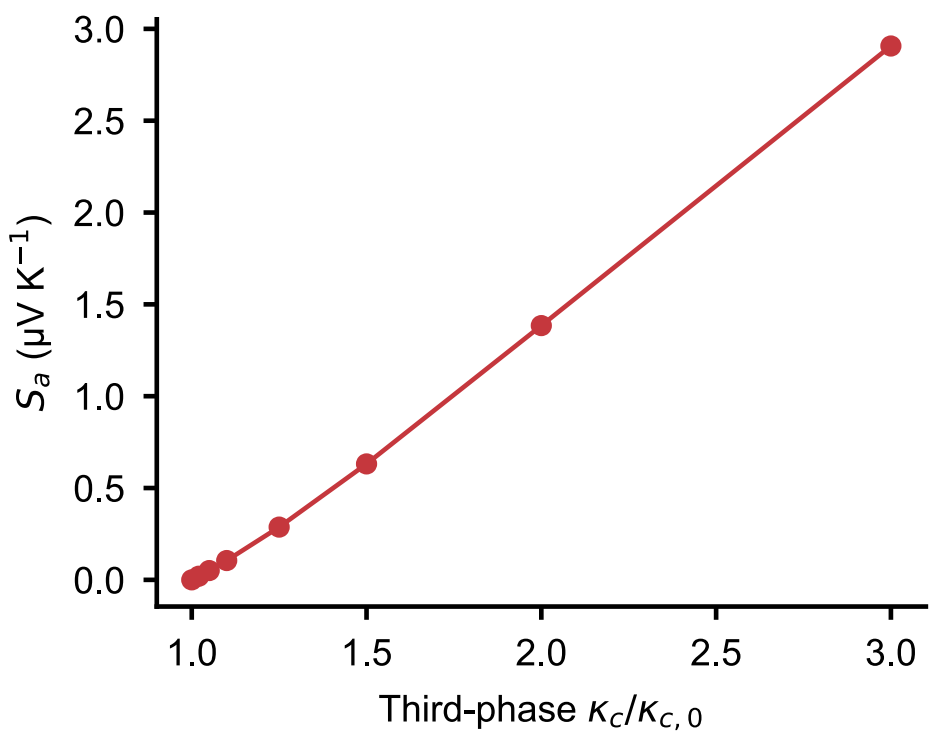

**Supplementary Fig. 2 | Material condition for the axial response.** The third phase of a pinwheel cell is first chosen as a linear combination of the other two ($\kappa_c = \kappa_{c,0}$), which gives $S_a = 0$. Increasing its thermal conductivity detunes the condition and $S_a$ grows.

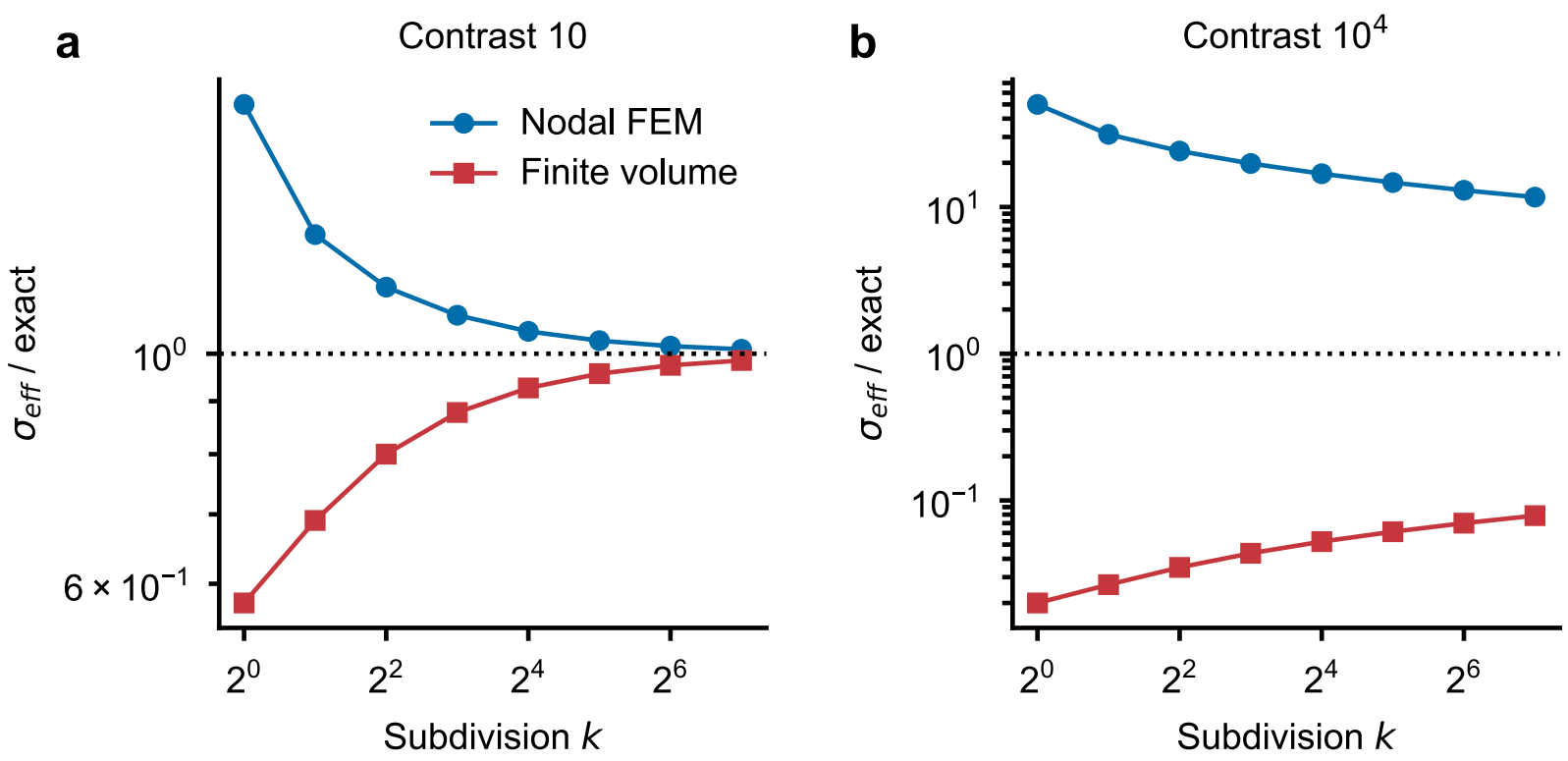

**Supplementary Fig. 3 | Convergence for the two-phase checkerboard.** Effective conductivity at contrasts of 10 and $10^4$ under uniform refinement, from nodal finite elements and finite volumes, compared with the exact value $\sqrt{(\sigma_1\sigma_2)}$.

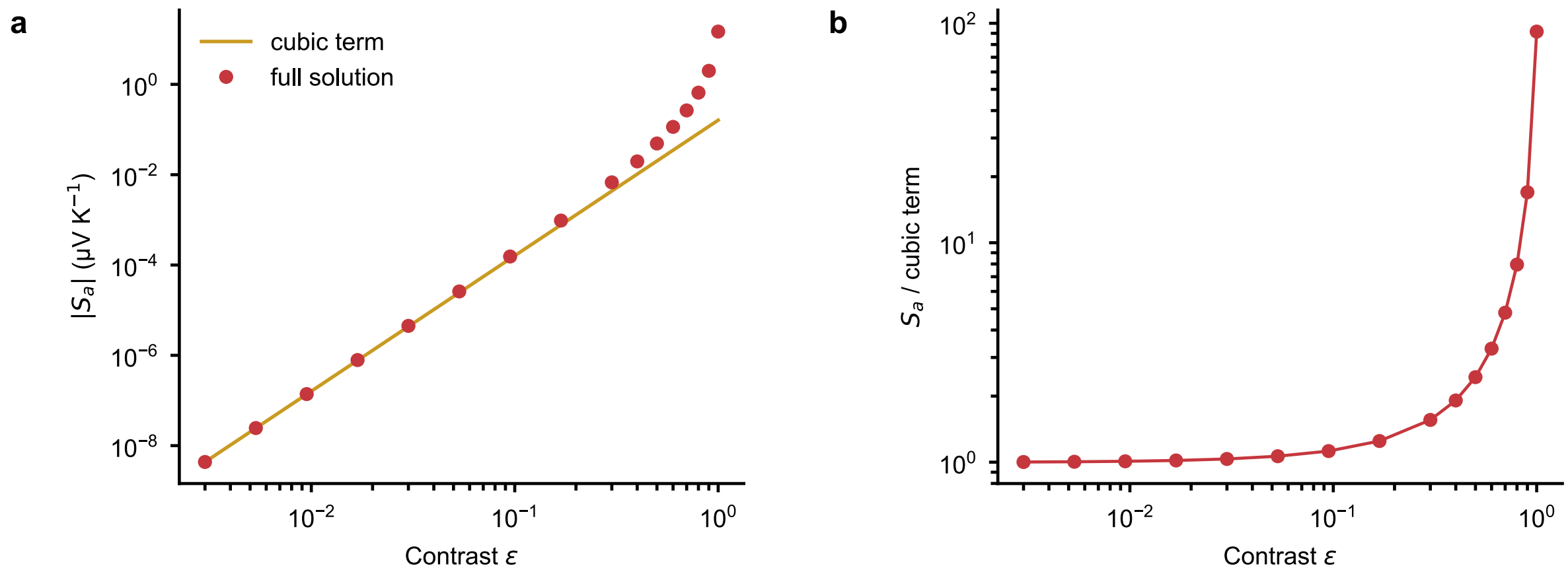

**Supplementary Fig. 4 | Cubic onset and full-contrast growth of the axial thermopower. a**, $|S_a|$ of the reference pinwheel (finite volumes, pixels refined fourfold) along the path $s_i(\varepsilon) = s_0 + \varepsilon(s_i - s_0)$ with $s_0 = (s_p + s_n)/2$, which reaches the real materials at $\varepsilon = 1$, and the cubic term of the expansion for this reference, $-0.1609\varepsilon^3$ μV K$^{-1}$, from Methods equation (9) with $R = s_0^{-1}$ and all three phase contrasts (equation (11) applies only to the phase-3 reference). **b**, Ratio of the full value to the cubic term. The ratio tends to 1 as $\varepsilon \to 0$ and grows to about 90 at $\varepsilon = 1$, where the series, convergent for this reference, is dominated by higher orders.

**Supplementary Table 1 | Interface resistance in the two-phase p4 pinwheel.**

| Cell size | $\rho_c$ ($\Omega$ m$^2$) | $R_K$ (m$^2$ K W$^{-1}$) | Two phases, $S_a$ (μV K$^{-1}$) | Mirror image (μV K$^{-1}$) | Three phases (μV K$^{-1}$) |
|---|---|---|---|---|---|
| 1 mm | 0 | 0 | 0.00 | 0.00 | −14.74 |
| 1 mm | $10^{-9}$ | 0 | −1.77 | 1.77 | −16.31 |
| 1 mm | $10^{-8}$ | 0 | −11.75 | 11.75 | −19.83 |
| 1 mm | $10^{-7}$ | 0 | −16.76 | 16.76 | −16.48 |
| 1 mm | $10^{-6}$ | 0 | −3.93 | 3.93 | −13.14 |
| 1 mm | 0 | $10^{-5}$ | 0.38 | −0.38 | −14.56 |
| 1 mm | $10^{-9}$ | $10^{-5}$ | −1.33 | 1.33 | −16.13 |
| 1 mm | $10^{-8}$ | $10^{-5}$ | −11.17 | 11.17 | −19.62 |
| 1 mm | $10^{-7}$ | $10^{-5}$ | −16.39 | 16.39 | −16.31 |
| 1 mm | $10^{-6}$ | $10^{-5}$ | −3.86 | 3.86 | −13.01 |
| 0.3 mm | 0 | 0 | 0.00 | 0.00 | −14.74 |
| 0.3 mm | $10^{-9}$ | 0 | −5.35 | 5.35 | −18.04 |
| 0.3 mm | $10^{-8}$ | 0 | −18.19 | 18.19 | −19.64 |
| 0.3 mm | $10^{-7}$ | 0 | −9.28 | 9.28 | −13.35 |
| 0.3 mm | $10^{-6}$ | 0 | −1.30 | 1.30 | −15.33 |
| 0.3 mm | 0 | $10^{-5}$ | 1.18 | −1.18 | −14.15 |
| 0.3 mm | $10^{-9}$ | $10^{-5}$ | −3.75 | 3.75 | −17.33 |
| 0.3 mm | $10^{-8}$ | $10^{-5}$ | −16.41 | 16.41 | −18.82 |
| 0.3 mm | $10^{-7}$ | $10^{-5}$ | −8.67 | 8.67 | −12.70 |
| 0.3 mm | $10^{-6}$ | $10^{-5}$ | −1.22 | 1.22 | −14.64 |

Pinwheel cells of Fig. 2a (two phases) and Fig. 2b (three phases), phases as in Fig. 2; zero-thickness neutral interfaces on every unlike-phase face as in Supplementary Table 11; finite-volume point estimates on the 24 × 24 raster refined four-fold (two-fold refinement changes $S_a$ by at most 2.6%). With ideal interfaces the two-phase cell gives $S_a = 0$.

**Supplementary Table 2 | Axial coefficient of two-phase cells without three-fold rotation.**

| Cell | Symmetry | $S_a$ (μV $K^{-1}$) |
|---|---|---|
| Hierarchical laminate, p-type/low-σ n-type | p2 | −1.08 |
| Rod and block cell, p-type/metal | p1 | −0.47 to −0.35 |
| Pinwheel cells of Fig. 2a (p4) | p4 | 0 |

Rod and block cell: p-type phase $\sigma = 10^5$ S $m^{-1}$, S = 200 μV $K^{-1}$, $\kappa = 1.5$ W $m^{-1}$ $K^{-1}$; metal $10^7$ S $m^{-1}$, 2 μV $K^{-1}$, 50 W $m^{-1}$ $K^{-1}$; certified interval for the fixed 64 × 64 raster, whose principal axes of $\sigma_{eff}$ and $L_{eff}$ differ by 1.9°.

## Supplementary Table 3 | Certified continuum intervals.

| Cell | $S_a$ (μV K$^{-1}$) | $S_{xx}$ (μV K$^{-1}$) | $\sigma_{eff}$ ($10^4$ S m$^{-1}$) | $\lvert S_a \rvert$/half-span |
|---|---|---|---|---|
| Pinwheel p/n/weak (Fig. 2b) | −15.65 to −15.43 | −4.20 to −3.97 | 1.662 to 1.664 | 0.077 to 0.079 |
| Pinwheel p/n/Cu | 0.47 to 0.64 | 37.57 to 37.74 | 48.299 to 48.330 | 0.002 to 0.004 |
| Design A (64 × 64 raster) | 49.52 to 50.15 | 8.16 to 8.79 | 3.219 to 3.225 | 0.260 to 0.264 |
| Design B (64 × 64 raster) | −14.63 to −13.44 | 12.68 to 13.87 | 0.536 to 0.538 | 0.070 to 0.077 |
| Design C, p/n/carbon | −55.48 to −55.19 | −25.43 to −25.14 | 0.646 to 0.648 | 0.290 to 0.292 |
| Design D, p/n/carbon/Cu | −61.41 to −61.11 | −31.91 to −31.61 | 0.750 to 0.752 | 0.321 to 0.324 |

Intervals follow from upper and lower bounds on the effective energy matrix. Half-span is (max $S_i$ − min $S_i$)/2 of the phases present; it is 200 μV K$^{-1}$ for the cells with the Fig. 2 pair and 190 μV K$^{-1}$ for cells with the p-type and n-type phases of Supplementary Table 14.

## Supplementary Table 4 | Fabrication tolerance of design A.

| Modification | Type | $S_a$ (μV K$^{-1}$) |
|---|---|---|
| Junction separation 0% | C | 43.8 to 54.4 |
| Junction separation 1% | C | 46.7 to 57.2 |
| Junction separation 2% | C | 46.7 to 57.2 |
| Junction separation 4% | C | 46.6 to 57.1 |
| Corner rounding 1% | C | 47.4 to 58.1 |
| Corner rounding 3% | C | 48.9 to 59.8 |
| Corner rounding 5% | C | 50.2 to 61.4 |
| Random ±2% boundary error, p4 kept (24 samples) | P | 52.1 ± 0.8 (min 49.0) |
| Random ±2% boundary error, symmetry free (24 samples) | P | 51.2 ± 4.6 (min 29.7) |
| Same, 4% separation protected (24 samples) | P | 52.1 ± 0.3 (min 51.4) |

C, certified interval on the fixed raster; P, finite-volume point estimate (mean ± sample standard deviation over the ensemble).

## Supplementary Table 5 | Winding-number voltages in a thermally poor matrix.

| n | Corrected voltage (μV) | $nS_aq'/\kappa_{eff}$ (μV) | $q'$ (W m$^{-1}$) |
|---|---|---|---|
| 1 | 6.70 | 6.69 | 0.1667 |
| 2 | 13.94 | 13.91 | 0.1733 |
| 3 | 22.08 | 21.96 | 0.1824 |

Matrix thermal conductivity 3% of that of the strip; $q'$ is the computed heat input per unit length at the line source. Values on a 1024 × 1024 grid (6.8 × $10^5$ elements); the deviation from $nS_aq'/\kappa_{eff}$ decreases from 3.4–4.0% on a 256 × 256 grid and 1.3–1.6% on a 512 × 512 grid.

**Supplementary Table 6 | Colour-group prediction across three-dimensional structures.**

| Family | Structures | Axial z allowed | Max $\lvert S_{a,z}\rvert$, allowed (μV K$^{-1}$) | Max $\lvert S_{a,z}\rvert$, forbidden (μV K$^{-1}$) |
|---|---|---|---|---|
| Segmented woodpiles | 82 | 81 | 45.7 | 2e-8 |
| Straight woodpiles | 17 | 0 | — | 1e-11 |
| RCSR-derived nets | 116 | 49 | 53.3 | 6e-9 |

Finite-volume values on 16- or 24-voxel cells. In every allowed case the finite-volume estimate was non-zero; the smallest values are unresolved (Supplementary Table 16).

**Supplementary Table 7 | Recolouring that restores the parent symmetry.**

| Net | Parent space group | Handed rule: colour point group | Handed rule: $S_{a,z}$ (μV $K^{-1}$) | Folded rule: colour point group | Folded rule: $S_{a,z}$ (μV $K^{-1}$) |
|---|---|---|---|---|---|
| bbr | P4122 | 4 | 13.06 | 422 | $\|S_{a,z}\| < 10^{-10}$ |
| boq | P-42m | −4 | −3.04 | −42m | $\|S_{a,z}\| < 10^{-10}$ |
| noo | P4/mnc | 4/m | −4.13 | 4/mmm | $\|S_{a,z}\| < 10^{-11}$ |
| nou | P4/mbm | 4/m | 5.99 | 4/mmm | $\|S_{a,z}\| < 10^{-10}$ |
| pll | P4122 | 4 | −5.74 | 422 | $\|S_{a,z}\| < 10^{-10}$ |
| svs | P4/mcc | 4/m | 5.89 | 4/mmm | $\|S_{a,z}\| < 10^{-10}$ |
| tfv | P-4n2 | −4 | 11.31 | −42m | $\|S_{a,z}\| < 10^{-10}$ |
| unm | P42212 | 4 | −16.49 | 422 | $\|S_{a,z}\| < 10^{-10}$ |
| uoj | P-4n2 | −4 | −3.75 | −42m | $\|S_{a,z}\| < 10^{-10}$ |
| uov | P-4c2 | −4 | 2.47 | −42m | $\|S_{a,z}\| < 10^{-10}$ |
| upa | P42212 | 4 | −16.72 | 422 | $\|S_{a,z}\| < 10^{-10}$ |
| wft | P-421c | −4 | −7.54 | −42m | $\|S_{a,z}\| < 10^{-10}$ |
| wfu | P-421c | −4 | −9.65 | −42m | $\|S_{a,z}\| < 10^{-11}$ |
| wgu | P42/mbc | 4/m | −9.34 | 4/mmm | $\|S_{a,z}\| < 10^{-11}$ |
| whc | P-4c2 | −4 | −1.28 | −42m | $\|S_{a,z}\| < 10^{-10}$ |
| xft | I41md | 4 | 24.89 | 4mm | $\|S_{a,z}\| < 10^{-10}$ |
| xfw | P-4c2 | −4 | 5.56 | −42m | $\|S_{a,z}\| < 10^{-10}$ |
| xfy | P-4n2 | −4 | 1.29 | −42m | $\|S_{a,z}\| < 10^{-11}$ |
| yfg | P4122 | 4 | −12.53 | 422 | $\|S_{a,z}\| < 10^{-11}$ |
| zfa | P42212 | 4 | 14.48 | 422 | $\|S_{a,z}\| < 10^{-11}$ |
| zfb | P42/mbc | 4/m | −15.42 | 4/mmm | $\|S_{a,z}\| < 10^{-11}$ |
| zfs | P4122 | 4 | 3.00 | 422 | $\|S_{a,z}\| < 10^{-11}$ |
| zfu | P41212 | 4 | −7.52 | 422 | $\|S_{a,z}\| < 10^{-10}$ |
| zlg | P42/mmc | 4/m | −6.19 | 4/mmm | $\|S_{a,z}\| < 10^{-10}$ |

These 24 nets remain three-phase but change from allowed to forbidden (handed-rule data: Supplementary Table 16). With $q = \theta \bmod 90°$ and $\varphi = \min(q, 90° - q)$, struts are n-type for $22.5° < \varphi < 45°$ and p-type otherwise, including vertical struts and ties at 0°, 22.5° and 45° ($10^{-6}$° tolerance); p-type takes priority at overlaps. The rule preserves the parent symmetry within 4/mmm, including mirrors containing the axis and perpendicular two-fold axes; eight of the 24 parent groups are 422, without mirrors. These parent groups forbid $S_{a,z}$. Geometry, materials and voxel grid are unchanged; the p/n labels and fractions change. Values are 16-voxel finite-volume residuals; symmetry makes the continuum value zero.

**Supplementary Table 8 | Certified axial coefficients of woodpiles and net-derived structures.**

| Structure | Colour group | FEM, k = 1, 2, 3 (μV $K^{-1}$) | FV, k = 1, 2, 3 (μV $K^{-1}$) | Certified $S_{a,z}$ (μV $K^{-1}$) |
|---|---|---|---|---|
| Woodpile W1 | 4 | 33.0, 34.7, 35.4 | 45.0, 43.9, 43.6 | 22.7 to 56.0 |
| Woodpile W2 | 4 | 54.3, 60.4, 63.0 | 82.3, 81.1, 80.7 | 57.4 to 86.1 |
| cyc-derived | 4/m | 52.2, 52.2, 52.3 | 53.3, 53.1, 52.9 | 47.7 to 57.5 |
| unm-derived | 4 | −12.1, −12.4, −12.4 | −13.1, −12.8, −12.7 | −15.4 to −9.7 |
| bbr-derived | 4 | 13.7, 13.5, 13.5 | 13.1, 13.4, 13.5 | 6.5 to 20.6 |
| Straight-rod woodpile | 422 | 0.0, 0.0, 0.0 | 0.0, 0.0, 0.0 | 0 (symmetry) |

Fixed voxel geometries refined k-fold: woodpile W1 16 and W2 24 cubic voxels per edge; nets 16 cubic voxels along a, which sets c/a to 1.5 (cyc, unm) and 0.75 (bbr); Supplementary Table 9 restores the original c/a. Mirror images reverse every entry.

**Supplementary Table 9 | Woodpiles with regularized junctions and nets with original c/a ratios.**

| Structure | Changed volume (%) | FEM, k = 1, 2, 3 (μV $K^{-1}$) | FV, k = 1, 2, 3 (μV $K^{-1}$) | Certified $S_{a,z}$ (μV $K^{-1}$) |
|---|---|---|---|---|
| W1, fused | 0.68 | 30.2, 30.8, 31.0 | 33.2, 32.2, 31.8 | 25.3 to 37.3 |
| W1, separated | 1.27 | 40.1, 41.2, 41.5 | 44.9, 43.5, 43.0 | 35.6 to 48.7 |
| W2, fused | 0.38 | 61.7, 63.0, 63.3 | 67.8, 65.7, 65.0 | 60.2 to 67.6 |
| W2, separated | 0.55 | 78.4, 79.3, 79.6 | 82.7, 81.4, 80.9 | 76.6 to 83.7 |
| cyc-derived, c/a = 1.414 | — | 52.3, 52.3 | 53.5, 53.2 | 44.4 to 61.1 |
| unm-derived, c/a = 1.468 | — | −12.1, −12.4 | −13.1, −12.8 | −17.4 to −7.6 |
| bbr-derived, c/a = 0.770 | — | 13.8, 13.6 | 13.1, 13.4 | 2.9 to 24.3 |

Fused junctions fill weak voxels next to like-phase edge contacts; separated junctions replace the contacting voxels by the weak phase; W2 variants also clear four weak/n/weak/p crossings. Nets are refined to k = 2. Certified intervals are rounded outward.

**Supplementary Table 10 | Three-dimensional structures with an electrically insulating matrix.**

| Structure | Weak matrix, k = 2 (μV $K^{-1}$) | Insulator, κ = 1.0 (μV $K^{-1}$) | Insulator, κ = 0.3 (μV $K^{-1}$) | p–n face percolation |
|---|---|---|---|---|
| W2, separated | 74.3 to 86.0 | 75.3 to 87.4 (k = 2) | 61.0 to 71.6 (k = 2) | x, y, z |
| W1, separated | 31.4 to 52.9 | 31.4 to 53.8 (k = 2) | 26.7 to 46.4 (k = 2) | x, y, z |
| cyc-derived | 44.4 to 61.1 | 44.7 to 61.8 (k = 2) | 36.8 to 56.2 (k = 2) | x, y, z |
| unm-derived | −17.4 to −7.6 | −17.5 to −7.6 (k = 2) | −16.6 to −4.7 (k = 2) | x, y, z |
| bbr-derived | 2.9 to 24.3 | 3.4 to 24.8 (k = 2) | 1.9 to 14.8 (k = 4) | x, y, z |

Certified $S_{a,z}$ intervals (rounded outward) on the stated voxel geometries refined k-fold; the insulator has $\sigma = 10^{-6}$ S $m^{-1}$, S = 0 and κ in W $m^{-1}$ $K^{-1}$. Percolation: directions in which the union of p-type and n-type voxels connects through shared faces across the periodic cell (exact graph test).

## Supplementary Table 11 | Effect of interface resistance on design A.

| Cell size | $\rho_c$ ($\Omega$ m$^2$) | $S_a$ ($\mu$V K$^{-1}$) | $\sigma_{eff}$ ($10^4$ S m$^{-1}$) | $PF_a$ ($\mu$W m$^{-1}$ K$^{-2}$) |
|---|---|---|---|---|
| 1 mm | 0 | 52.0 | 3.17 | 85.8 |
| 1 mm | 1e-09 | 54.2 | 2.85 | 83.5 |
| 1 mm | 1e-08 | 58.1 | 1.57 | 53.1 |
| 1 mm | 1e-07 | 58.4 | 0.30 | 10.2 |
| 1 mm | 1e-06 | 51.3 | 0.04 | 1.0 |
| 0.3 mm | 0 | 52.0 | 3.17 | 85.8 |
| 0.3 mm | 1e-09 | 56.3 | 2.34 | 74.3 |
| 0.3 mm | 1e-08 | 58.9 | 0.74 | 25.7 |
| 0.3 mm | 1e-07 | 56.0 | 0.10 | 3.2 |
| 0.3 mm | 1e-06 | 42.9 | 0.01 | 0.3 |

Zero-thickness interfaces with Kapitza resistance $R_K = 0$; finite-volume point estimates. With $R_K = 10^{-5}$ m$^2$ K W$^{-1}$, $S_a$ changes by less than 10% and $PF_a$ by up to 15%.

**Supplementary Table 12 | Orthogonal-gradient measurement on square specimens.**

| Structure | N | Probes and termination | $S_{xy}$ (µV K$^{-1}$) | $S_{yx}$ (µV K$^{-1}$) | Half-difference (µV K$^{-1}$) | Half-sum (µV K$^{-1}$) | Half-difference / bulk (%) |
|---|---|---|---|---|---|---|---|
| Design A, junctions separated by 4% | 4 | mid-height probes | 50.11 | −50.11 | 50.11 | 0.00 | 96.0–97.3 |
| Design A, junctions separated by 4% | 4 | both probes tilted, both directions | 38.94 | −61.14 | 50.04 | −11.10 | 95.9–97.1 |
| Design A, junctions separated by 4% | 4 | probes tilted, y drive only | 38.94 | −50.11 | 44.53 | −5.59 | 85.3–86.4 |
| Design A, junctions separated by 4% | 4 | probes tilted oppositely | 38.94 | −38.94 | 38.94 | 0.00 | 74.6–75.6 |
| Design A, junctions separated by 4% | 4 | both probes shifted | 43.69 | −43.69 | 43.69 | 0.00 | 83.7–84.8 |
| Design A, junctions separated by 4% | 8 | mid-height probes | 50.90 | −50.90 | 50.90 | 0.00 | 97.5–98.8 |
| Design A, junctions separated by 4% | 8 | both probes tilted, both directions | 45.30 | −56.42 | 50.86 | −5.56 | 97.4–98.7 |
| Design A, junctions separated by 4% | 8 | probes tilted, y drive only | 45.30 | −50.90 | 48.10 | −2.80 | 92.1–93.3 |
| Design A, junctions separated by 4% | 8 | probes tilted oppositely | 45.30 | −45.30 | 45.30 | 0.00 | 86.8–87.9 |
| Design A, junctions separated by 4% | 8 | both probes shifted | 47.69 | −47.69 | 47.69 | 0.00 | 91.3–92.5 |
| Design A, junctions separated by 4% | 8 | mid-height probes, shifted termination | 52.33 | −47.24 | 49.79 | 2.55 | 95.4–96.6 |
| Design A, junctions separated by 4% | 8 | both probes tilted, both directions, shifted termination | 42.28 | −45.07 | 43.68 | −1.39 | 83.7–84.8 |
| Design A, junctions separated by 4% | 8 | probes tilted, y drive only, shifted termination | 42.28 | −47.24 | 44.76 | −2.48 | 85.7–86.9 |
| Design A, junctions separated by 4% | 8 | probes tilted oppositely, shifted termination | 42.28 | −46.27 | 44.28 | −1.99 | 84.8–85.9 |
| Design A, junctions separated by 4% | 8 | both probes shifted, shifted termination | 49.51 | −50.05 | 49.78 | −0.27 | 95.4–96.6 |
| Pinwheel of Fig. 2b | 4 | mid-height probes | −11.26 | 11.26 | −11.26 | 0.00 | 71.9–73.0 |
| Pinwheel of Fig. | 4 | both probes tilted, both | −12.24 | 12.60 | −12.42 | 0.18 | 79.3–80.5 |

| Structure | N | Probes and termination | $S_{xy}$ (μV $K^{-1}$) | $S_{yx}$ (μV $K^{-1}$) | Half-difference (μV $K^{-1}$) | Half-sum (μV $K^{-1}$) | Half-difference / bulk (%) |
|---|---|---|---|---|---|---|---|
| 2b | | directions | | | | | |
| Pinwheel of Fig. 2b | 4 | probes tilted, y drive only | −12.24 | 11.26 | −11.75 | −0.49 | 75.1–76.2 |
| Pinwheel of Fig. 2b | 4 | probes tilted oppositely | −12.24 | 12.24 | −12.24 | 0.00 | 78.2–79.4 |
| Pinwheel of Fig. 2b | 4 | both probes shifted | −20.57 | 20.57 | −20.57 | 0.00 | 131.4–133.4 |
| Pinwheel of Fig. 2b | 8 | mid-height probes | −13.27 | 13.27 | −13.27 | 0.00 | 84.8–86.1 |
| Pinwheel of Fig. 2b | 8 | both probes tilted, both directions | −13.77 | 13.93 | −13.85 | 0.08 | 88.5–89.8 |
| Pinwheel of Fig. 2b | 8 | probes tilted, y drive only | −13.77 | 13.27 | −13.52 | −0.25 | 86.4–87.7 |
| Pinwheel of Fig. 2b | 8 | probes tilted oppositely | −13.77 | 13.77 | −13.77 | 0.00 | 88.0–89.3 |
| Pinwheel of Fig. 2b | 8 | both probes shifted | −18.09 | 18.09 | −18.09 | 0.00 | 115.6–117.3 |
| Pinwheel of Fig. 2b | 8 | mid-height probes, shifted termination | −15.23 | 16.63 | −15.93 | 0.70 | 101.8–103.3 |
| Pinwheel of Fig. 2b | 8 | both probes tilted, both directions, shifted termination | −23.88 | 10.14 | −17.01 | −6.87 | 108.7–110.3 |
| Pinwheel of Fig. 2b | 8 | probes tilted, y drive only, shifted termination | −23.88 | 16.63 | −20.25 | −3.63 | 129.4–131.3 |
| Pinwheel of Fig. 2b | 8 | probes tilted oppositely, shifted termination | −23.88 | 16.26 | −20.07 | −3.81 | 128.2–130.1 |
| Pinwheel of Fig. 2b | 8 | both probes shifted, shifted termination | −21.56 | 12.00 | −16.78 | −4.78 | 107.2–108.8 |
| Design A, junctions separated by 4% | 16 | mid-height probes | 51.28 | −51.28 | 51.28 | 0.00 | 98.2–99.5 |
| Design A, junctions separated by 4% | 16 | both probes tilted, both directions | 48.48 | −54.04 | 51.26 | −2.78 | 98.2–99.5 |
| Design A, junctions separated by 4% | 16 | probes tilted, y drive only | 48.48 | −51.28 | 49.88 | −1.40 | 95.6–96.8 |
| Design A, junctions separated by 4% | 16 | probes tilted oppositely | 48.48 | −48.48 | 48.48 | 0.00 | 92.9–94.1 |
| Design A, junctions separated by 4% | 16 | both probes shifted | 49.68 | −49.68 | 49.68 | 0.00 | 95.2–96.4 |
| Pinwheel of Fig. 2b | 16 | mid-height probes | −14.37 | 14.37 | −14.37 | 0.00 | 91.8–93.2 |
| Pinwheel of Fig. 2b | 16 | both probes tilted, both directions | −14.62 | 14.70 | −14.66 | 0.04 | 93.7–95.1 |

| Structure | N | Probes and termination | $S_{xy}$ (µV $K^{-1}$) | $S_{yx}$ (µV $K^{-1}$) | Half-difference (µV $K^{-1}$) | Half-sum (µV $K^{-1}$) | Half-difference / bulk (%) |
|---|---|---|---|---|---|---|---|
| Pinwheel of Fig. 2b | 16 | probes tilted, y drive only | −14.62 | 14.37 | −14.50 | −0.13 | 92.6–94.0 |
| Pinwheel of Fig. 2b | 16 | probes tilted oppositely | −14.62 | 14.62 | −14.62 | 0.00 | 93.4–94.8 |
| Pinwheel of Fig. 2b | 16 | both probes shifted | −16.82 | 16.82 | −16.82 | 0.00 | 107.4–109.0 |

Finite-element point estimates (P) on square specimens of N × N cells. Each specimen is driven once along x and once along y; the hot and cold faces are isothermal ($\Delta T = 1$ K), the other faces are adiabatic and all faces are electrically insulating, and point probes draw no current. The bulk reference is the certified interval of the same raster. Tilted probes are displaced by ±1/8 cell along the drive axis; shifted probes by 1/4 cell; a shifted termination translates the pattern by 1/4 cell before cutting the specimen.

## Supplementary Table 13 | Transverse voltage of finite samples.

| Sample | $V_\perp$ (μV) | $V_\perp/(S_a \Delta T\ W/H)$ |
|---|---|---|
| Design A as drawn (point contacts), 8 × 4 cells | 40.0 | 0.40 |
| Design A as drawn (point contacts), 16 × 8 cells | 35.7 | 0.36 |
| Design A, junctions separated by 4%, 8 × 4 cells | 102.2 | 0.989 |
| Design A, junctions separated by 4%, 16 × 8 cells | 102.8 | 0.995 |
| Homogeneous effective medium, 8 × 4 | 99.7 | 1.000 |
| Two-phase p4 control, 8 × 4 | 2.93 | 0.029 |
| Mirror-symmetric control, 8 × 4 | 8e-10 | 0 |

$\Delta T$ = 1 K between top and bottom, W/H = 2, probes at mid-height on both sides; finite-element point estimates. For separated junctions the ratio uses the bulk $S_a$ on the same grid. For the design as drawn it uses the continuum $S_a$; the low ratio reflects point contacts that the grid does not resolve (the same grid gives a bulk $S_a$ of 15.8 μV $K^{-1}$, because without resolving their singular fields the discretization couples the point contacts like finite junctions). The two-phase control assumes ideal interfaces (Supplementary Table 1).

**Supplementary Table 14 | Material parameters at 300 K.**

| Phase | σ (S m$^{-1}$) | S (μV K$^{-1}$) | κ (W m$^{-1}$ K$^{-1}$) | Role |
|---|---|---|---|---|
| p | 1.0e+5 | 200.0 | 1.20 | p-type BiSbTe |
| n | 1.0e+5 | −180.0 | 1.30 | n-type $Bi_2Te_{2.7}Se_{0.3}$ |
| Cu | 6.0e+7 | 1.8 | 400.00 | metal |
| Ni | 1.4e+7 | −19.0 | 90.00 | metal |
| PEDOT | 1.0e+5 | 20.0 | 0.30 | polymer conductor (search pool only) |
| carbon | 1.0e+3 | 10.0 | 0.30 | carbon composite |
| polymer_void | 1.0e-6 | 0.0 | 0.20 | insulating polymer (search pool only) |
| p, n (Fig. 2) | 1.0e+5 | ±200.0 | 1.00 | reference thermoelectric pair |
| weak (Fig. 2) | 1.0e+3 | 0.0 | 0.20 | reference weak phase |
| third (design A) | 10.45 | 10.0 | 0.9012 | passes heat, blocks charge |

## Supplementary Table 15 | Stacked and helical three-dimensional cells.

| Cell | Point group | $S_{a,z}$, FEM / FV, k = 3 (μV K⁻¹) | Certified $S_{a,z}$ (μV K⁻¹) |
|---|---|---|---|
| Staggered pinwheel layers | 4 | 4.90 / 17.56 | 1.92 to 20.06 |
| Three-phase $4_1$ helix | 4 | −4.48 / −3.98 | −8.26 to −0.26 |
| Opposite-handed helix | 4 | 4.48 / 3.98 | 0.26 to 8.26 |
| Two-phase helix, p/n | 4 | 0.00 / 0.00 | 0 (two-phase) |
| Two-phase helix, p/weak | 4 | 0.00 / 0.00 | 0 (two-phase) |
| Glide-related stack | 4mm-type | 0.00 / 0.00 | 0 (symmetry) |

Fixed 16-voxel geometries refined threefold. Certified intervals from the conforming upper bound and the complementary lower bound; zero entries follow exactly from the two-phase reduction under the four-fold axis (Supplementary Note 2) or from the glide symmetry.

**Supplementary Table 16 | All 116 computed RCSR-derived structures.**

| Net | Parent space group | Colour point group | $S_{a,z}$ (μV K$^{-1}$) | Net | Parent space group | Colour point group | $S_{a,z}$ (μV K$^{-1}$) |
|---|---|---|---|---|---|---|---|
| asd | P4/mmm | 4/m | 12.28 | sfo | I4/mmm | 4/mmm | forbidden |
| bbr | P4122 | 4 | 13.06 | sne | I-42d | −4 | 4.18 |
| bct | I4/mmm | 4/mmm | forbidden | snp | P42/mmc | 4/mmm | forbidden |
| bcu-c | I41/amd | 4/mmm | forbidden | sql-c | P42/mmc | 4/mmm | forbidden |
| boq | P-42m | −4 | −3.04 | sql-c*-w | P4/ncc | 4/mmm | forbidden |
| bsa | I4/mmm | 4/mmm | forbidden | srs-c* | P4222 | 422 | forbidden |
| bsd | P4/mmm | 4/mmm | forbidden | sub | P4/mmm | 4/mmm | forbidden |
| bsn | I41/amd | 4/mmm | forbidden | sup | P4/mmm | 4/mmm | forbidden |
| cap | I41/amd | 4/mmm | forbidden | sve | P4/mbm | 4/m | −2.69 |
| cds-a | P42/mmc | 4/mmm | forbidden | svf | I4/mcm | 4/mmm | forbidden |
| cds-b | P42/mnm | 4/mmm | forbidden | svg | I4/mmm | 4/mmm | forbidden |
| crb | I4/mmm | 4/mmm | forbidden | svi | I4/mcm | 4/m | 3.98 |
| crb-b | P42/mnm | 4/mmm | forbidden | svi-x | I4/mcm | 4/m | −1.09 |
| cut | P4/mmm | 4/mmm | forbidden | svs | P4/mcc | 4/m | 5.89 |
| cyc | P42/mmc | 4/m | 53.35 | tfn | P42/mnm | 4/m | 6.87 |
| cyg | P4/mmm | 4/m | 12.21 | tfv | P-4n2 | −4 | 11.31 |
| ddi | I41/amd | 4/mmm | forbidden | ths-c3 | I41/amd | 4/mmm | forbidden |
| deh-d | I41/amd | 4/mmm | forbidden | trn | P42/mmc | 4/mmm | forbidden |
| dft | P42/mmc | 4/mmm | forbidden | tty | P4/mmm | 4/mmm | forbidden |
| dia-b4* | P-42m | 4/mmm | forbidden | ukg | P42/n | 4/m | −1.57 |
| dia-w | I41/amd | 4/mmm | forbidden | umw | P4122 | 422 | forbidden |
| dia-w-c | P42/nnm | 4/mmm | forbidden | umx | P42/mcm | 4/m | 5.17 |
| dia-w* | I4122 | 422 | forbidden | unc | P4122 | 422 | forbidden |
| dmd | P42/mmc | 4/mmm | forbidden | unc-c | I4122 | 422 | forbidden |
| dnq | P-4m2 | −4 | 7.71 | unm | P42212 | 4 | −16.49 |
| eci | P41212 | 4 | −4.13 | uoj | P-4n2 | −4 | −3.75 |
| edi | P-4m2 | −42m | forbidden | uov | P-4c2 | −4 | 2.47 |
| eft | P4/mmm | 4/mmm | forbidden | upa | P42212 | 4 | −16.72 |
| ela | P42/mnm | 4/m | −0.73 | utk | P4122 | 422 | forbidden |
| fnx | P4122 | 422 | forbidden | wfd | I4/m | 4/m | −2.05 |
| fsc | P4/mmm | 4/mmm | forbidden | wfe | P42/n | 4/m | −6.89 |
| fsd | P4/mmm | 4/mmm | forbidden | wff | I-42m | −4 | 2.84 |
| fse | I-4m2 | −42m | forbidden | wfs | I-42m | −4 | −22.73 |
| fsj | P4/mmm | 4/mmm | forbidden | wft | P-421c | −4 | −7.54 |
| fuq | I4/mmm | 4/mmm | forbidden | wfu | P-421c | −4 | −9.65 |
| fwl | P4/mmm | 4/mmm | forbidden | wfy | I-4 | −4 | −3.62 |
| fwq | P4/mmm | 4/mmm | forbidden | wgu | P42/mbc | 4/m | −9.34 |
| fws | P4/mmm | 4/mmm | forbidden | wgv | P42/mnm | 4/m | −9.52 |
| hcb-c | I4/mcm | 4/mmm | forbidden | wgw | P42/ncm | 4/m | −3.52 |
| lig-c | P42/nnm | 4/mmm | forbidden | wha | P-421c | −4 | −5.38 |

| Net | Parent space group | Colour point group | $S_{a,z}$ (μV $K^{-1}$) | Net | Parent space group | Colour point group | $S_{a,z}$ (μV $K^{-1}$) |
|---|---|---|---|---|---|---|---|
| lsz | P42/nnm | 4/mmm | forbidden | whb | P42/nmc | 4/mmm | forbidden |
| lvt | I41/amd | 4/mmm | forbidden | whc | P-4c2 | −4 | −1.28 |
| lvt-c | P42/nnm | 4/mmm | forbidden | xah | P4/mmm | 4/mmm | forbidden |
| mot | P4/mmm | 4/mmm | forbidden | xal | P4/mmm | 4/mmm | forbidden |
| mot-e | P4/mmm | 4/mmm | forbidden | xft | I41md | 4 | 24.89 |
| msw | P42/nnm | 4/mmm | forbidden | xfu | P42/n | 4/m | 9.51 |
| ncd | P42/mnm | 4/m | −9.92 | xfv-z | I-42m | −42m | forbidden |
| ncf | I4/m | 4/m | −5.70 | xfw | P-4c2 | −4 | 5.56 |
| noo | P4/mnc | 4/m | −4.13 | xfx | I-4m2 | −42m | forbidden |
| nou | P4/mbm | 4/m | 5.99 | xfy | P-4n2 | −4 | 1.29 |
| pcs | P4/mmm | 4/mmm | forbidden | yfg | P4122 | 4 | −12.53 |
| pll | P4122 | 4 | −5.74 | yfx | I4/m | 4/m | 10.51 |
| psb | P42/mmc | 4/mmm | forbidden | zfa | P42212 | 4 | 14.48 |
| pts | P42/mmc | 4/mmm | forbidden | zfb | P42/mbc | 4/m | −15.42 |
| pts-x | P42/mmc | 4/mmm | forbidden | zfs | P4122 | 4 | 3.00 |
| rtl | P42/mnm | 4/mmm | forbidden | zfu | P41212 | 4 | −7.52 |
| scu | P4/mmm | 4/mmm | forbidden | zlg | P42/mmc | 4/m | −6.19 |
| scu-c | I4/mmm | 4/mmm | forbidden | zll | I4/mmm | 4/mmm | forbidden |

Parent space group: the undecorated RCSR net. Colour point group: the point operations that leave every phase in place after the strut colouring, which decides whether $S_{a,z}$ is allowed; colouring can lower a parent group such as P4/mmm to 4/m. Colour groups are those of the 16-voxel maps used in the calculations, found by testing every voxel translation. For 115 nets they equal the groups of the continuous coloured-strut structures, obtained from the space-group operations that map every strut onto a strut of the same colour; for dia-b4*, whose struts are all p-type, the voxel map has 4/mmm and the continuous structure −42m, and both forbid $S_{a,z}$. All 67 forbidden nets contain only p-type struts and are two-phase, and all 49 allowed nets are three-phase. This follows from the colouring rule: a strut off the 0° and 45° directions changes colour under every mirror containing z and every two-fold axis perpendicular to z, while the four-fold or −4 axis keeps it, so among these nets the colour verdict coincides with the phase count; three-phase forbidden controls include the 17 straight woodpiles, the one segmented woodpile that keeps the group 422 and the 24 recoloured nets of Supplementary Table 7. Selection: tetragonal parent, at most eight vertices in the conventional cell, $0.5 \leq c/a \leq 1.6$ and struts fixed by the edge centres (3,749 → 116; Supplementary Note 5). Finite-volume values on cubic voxels, 16 along a and $N_z = 16c/a$ rounded to the nearest integer and then up to a multiple of 4, so the computed c/a is a multiple of 0.25 and differs from the RCSR value by a factor between 0.96 and 1.38; the colour groups and verdicts do not depend on this stretch, and the certified net values restore the original c/a. Values below a few μV $K^{-1}$ are point estimates: for ukg (−1.6 μV $K^{-1}$) refined certified intervals contain zero. Struts of radius 0.08a; "forbidden" marks colour groups that forbid

$S_{a,z}$, where $|S_{a,z}| < 10^{-8}$ μV $K^{-1}$.

## Supplementary Table 17 | Three phases with equal σκ.

| Refinement m | Upper-trial $S_a$ (μV $K^{-1}$) | Certified $S_a$ (μV $K^{-1}$) |
|---|---|---|
| 2 | 0.0567 | −5.14 to 5.14 |
| 4 | 0.0143 | −1.28 to 1.28 |
| 8 | 0.0036 | −0.32 to 0.32 |

Pinwheel of Fig. 2b with the weak phase replaced by $\sigma = 10^3$ S $m^{-1}$, S = 0, $\kappa = 100$ W $m^{-1}$ $K^{-1}$; all three phases have $\sigma\kappa = 10^5$ W S $m^{-2}$ $K^{-1}$. Corner-enriched primal and dual bounds as in Supplementary Table 3.

**Supplementary Table 18 | Idealized three-phase materials that give $|S_a|$ above the largest constituent Seebeck coefficient.**

| Phase | σ (S m⁻¹) | S (μV K⁻¹) | κ (W m⁻¹ K⁻¹) |
|---|---|---|---|
| 1 | 9.75e+7 | 88.1 | 0.054 |
| 2 | 2.18e+5 | −88.0 | 0.057 |
| 3 | 1.21e+5 | −32.6 | 499.838 |

No Wiedemann–Franz constraint was imposed and these materials are not realizable; they only show that no bound by the constituent coefficients exists; the search box was $\sigma = 10^2$–$10^8$ S $m^{-1}$, $|S| \leq 200$ μV $K^{-1}$ and $\kappa = 0.05$–501 W $m^{-1}$ $K^{-1}$, and several optimal values lie on its edges. Continuum $S_a$ = 105.4 to 105.7 μV $K^{-1}$ against a largest constituent |S| of 88.1 μV $K^{-1}$, a ratio of 1.196 to 1.200. Geometry of design C.

**Supplementary References**